%% file: example_paper.tex
\documentclass{article}

\usepackage{microtype}
\usepackage{graphicx}
\usepackage{subcaption}
\usepackage{booktabs} 
\usepackage{longtable, booktabs}
\usepackage{hyperref}
\usepackage{url}
\usepackage{multirow}
\usepackage{graphicx}
\usepackage{booktabs}
\usepackage{makecell}
\usepackage{caption}
\usepackage{svg}
\usepackage{wrapfig}
\usepackage{hhline}
\usepackage{subcaption}  
\usepackage{enumitem}
\usepackage[most]{tcolorbox}
\usepackage{cite}
\usepackage{tabularx}
\usepackage{array}
\usepackage[table]{xcolor}
\usepackage{caption}

\newcolumntype{Y}{>{\centering\arraybackslash}X}
\usepackage{pgfplots}
\pgfplotsset{compat=1.18}
\usepackage{tikz}
\usepackage{placeins}

\usepackage{hyperref}

\usepackage[preprint]{icml2026}

\usepackage{amsmath}
\usepackage{amssymb}
\usepackage{mathtools}
\usepackage{amsthm}

\usepackage[capitalize,noabbrev]{cleveref}

\theoremstyle{plain}

\theoremstyle{definition}

\theoremstyle{remark}

\usepackage[textsize=tiny]{todonotes}

\icmltitlerunning{SAKE: Towards Editing Auditory Attribute Knowledge of Large Audio-Language Models}

\begin{document}

\twocolumn[
  \icmltitle{SAKE: Towards Editing Auditory Attribute Knowledge of Large Audio-Language Models}



  \icmlsetsymbol{equal}{*}

  \begin{icmlauthorlist}
    \icmlauthor{Chih-Kai Yang}{equal,yyy}
    \icmlauthor{Yen-Ting Piao}{equal,yyy}
    \icmlauthor{Tzu-Wen Hsu}{comp}
    \icmlauthor{Szu-Wei Fu}{sch}
    \icmlauthor{Zhehuai Chen}{sch}
    \icmlauthor{Ke-Han Lu}{yyy}
    \icmlauthor{Sung-Feng Huang}{sch}
    \icmlauthor{Chao-Han Huck Yang}{sch}
    \icmlauthor{Yu-Chiang Frank Wang}{sch}
    \icmlauthor{Yun-Nung Chen}{yyy}
    \icmlauthor{Hung-yi Lee}{yyy}
  \end{icmlauthorlist}

  \icmlaffiliation{yyy}{National Taiwan University}
  \icmlaffiliation{comp}{DouDou Capital}
  \icmlaffiliation{sch}{NVIDIA}

  \icmlcorrespondingauthor{Chih-Kai Yang}{chihkaiyang1124@gmail.com}
  \icmlcorrespondingauthor{Hung-yi Lee}{hungyilee@ntu.edu.tw}

  \icmlkeywords{Machine Learning, ICML}

  \vskip 0.3in
]



\printAffiliationsAndNotice{}  


\begin{abstract}
Knowledge editing enables targeted updates without retraining, but prior work focuses on textual or visual facts, leaving abstract auditory perceptual knowledge underexplored. We introduce SAKE, the first benchmark for editing perceptual auditory attribute knowledge in large audio-language models (LALMs), which requires modifying acoustic generalization rather than isolated facts. We evaluate eight diverse editing methods on three LALMs across reliability, generality, locality, and portability, under single and sequential edits. Results show that most methods enforce edits reliably but struggle with auditory generalization, intra-attribute locality, and multimodal knowledge propagation, and often exhibit forgetting or degeneration in sequential editing. Additionally, fine-tuning the modality connector emerges as a more robust and balanced baseline compared with directly editing the LLM backbones. SAKE reveals key limitations of current methods and provides a foundation for developing auditory-specific LALM editing techniques.

\end{abstract}


\section{Introduction}

Large language models (LLMs)~\citep{touvron2023llama, grattafiori2024llama, hurst2024gpt} have achieved remarkable success across a wide range of language tasks. As they scale and are increasingly deployed, knowledge editing~\citep{efk, mend, ike, unke} has become a key technique for efficiently updating specific model knowledge without full retraining while minimizing catastrophic forgetting. Knowledge editing can also be extended to bias mitigation~\citep{chen2024large} and personalization~\citep{lu2025mobiedit}.


Recent advances have extended LLMs to multimodal settings, including large vision-language models (LVLMs)~\citep{llava, li2023blip} and large audio-language models (LALMs)~\citep{lu2025desta25Audio, qwen2audio, speechcopilot, audioflamingo, hurst2024gpt, lin2025preliminary, Lu2025Developing, tang2024salmonn}. As multimodal models become more prevalent, it is increasingly important to understand whether and how knowledge editing can be applied beyond the textual domain. While recent benchmarks have explored visual knowledge editing in LVLMs~\citep{mmedit, vlkeb, mcmke}, knowledge editing in the auditory modality remains largely unexplored.


Most knowledge editing studies~\citep{efk, mend, rome} focus on factual knowledge expressed as declarative propositions, often formalized as subject-relation-object triplets~\cite{rome} (e.g., “Paris
is the capital of France”). Visual knowledge editing in LVLMs largely follows this paradigm by treating images as alternative grounding for entity-centric facts. In contrast, perceptual knowledge is qualitatively different. Editing perceptual knowledge is conceptually closer to modifying style-level understanding, such as writing style in the text domain or image style in the vision domain, than to updating individual facts, as it operates on high-level abstractions rather than isolated factual associations.

Auditory attribute knowledge is a form of perceptual knowledge that reflects how models conceptualize attributes such as speaker gender and emotion. These attributes arise from perceptual abstractions over signals with substantial intra-class variability across speakers, prosody, and recording conditions. Editing such knowledge, therefore, requires modifying how models generalize perceptual evidence, rather than changing a factual statement. Hence, it is unclear whether editing methods developed for factual knowledge can generalize to abstract perceptual concepts in auditory modalities.


\begin{figure*}[ht]
    \centering
    \includegraphics[width=\linewidth]{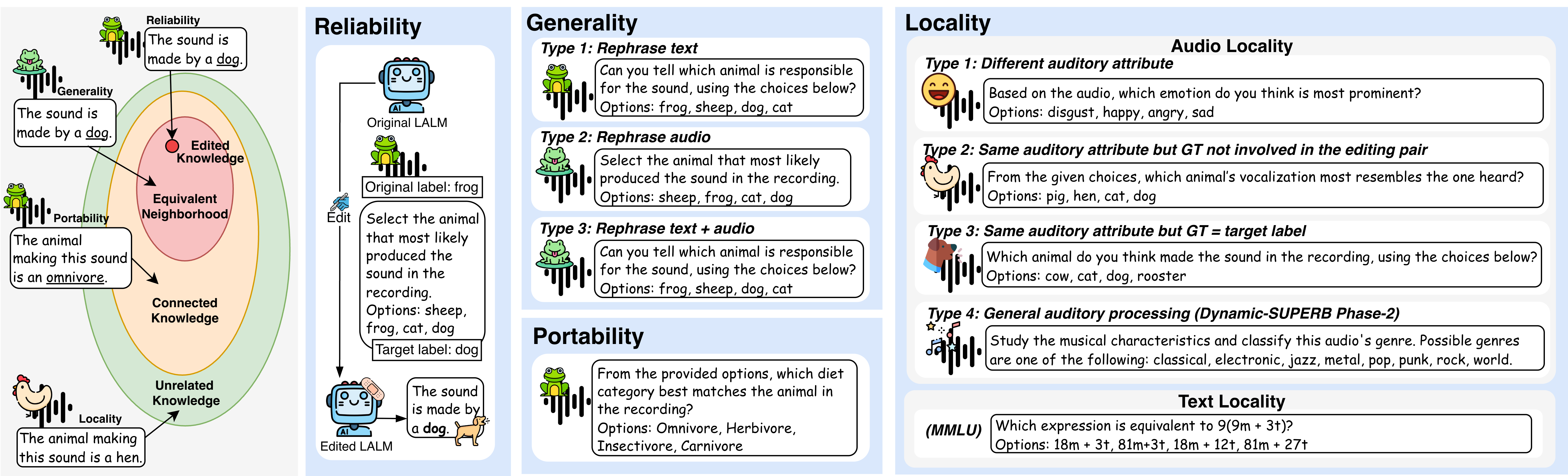}
    \vspace{-6mm}
    \caption{\textbf{Left:} Conceptual illustration of the knowledge scope affected by an edit. \textbf{Right:} Overview of SAKE benchmark, targeting auditory attributes including speaker gender, emotion, spoken language, and animal sounds. Reliability measures edit success, Generality its consistency across equivalent data, Locality the preservation of unrelated knowledge, and Portability its transfer to connected knowledge. For example, after editing ``frog'' to ``dog,'' the answer of the portability question should change from ``Insectivore'' to ``Omnivore.''}
    \label{fig:sake-overview}
\end{figure*}

To this end, we introduce \textbf{SAKE} (\textbf{S}peech and \textbf{A}udio Attribute \textbf{K}nowledge \textbf{E}diting Benchmark), the first benchmark for auditory attribute knowledge editing in LALMs (Figure~\ref{fig:sake-overview}). Auditory attribute knowledge editing targets modifications of an LALM’s perceptual understanding of specific attributes, such as changing its auditory knowledge of frogs’ vocalizations to that of dogs (Figure~\ref{fig:sake-overview}), while preserving related reasoning and unrelated knowledge. Such editing enables applications such as LALM error correction for misrecognition, debiasing, and voice assistant personalization, which are inherently more challenging than text-based settings due to the complexity of auditory modalities. SAKE spans diverse auditory attributes and evaluates editing methods along four dimensions~\citep{yao-etal-2023-editing}: reliability, generality, locality, and portability.


We evaluate eight representative editing methods on three LALMs, DeSTA2.5-Audio~\citep{lu2025desta25Audio}, Qwen2-Audio~\citep{qwen2audio}, and Audio Flamingo 3~\citep{af3}. While these methods have proven effective in textual domains, our results reveal persistent challenges in auditory attribute knowledge editing, including preserving intra-attribute knowledge and enabling edit propagation to connected reasoning. Under sequential editing, most methods suffer severe forgetting. Together, these findings highlight the unique challenges of auditory knowledge editing and the need for future advances in this research area.



Our contributions are three-fold: (1) We introduce SAKE, the first comprehensive benchmark for systematically evaluating auditory attribute knowledge editing of LALMs; (2) We show that existing methods struggle to preserve non-target auditory knowledge and to propagate edits during reasoning, revealing key limitations in editing abstract perceptual attributes; (3) Through extensive analysis, we show that fine-tuning the modality connector is strong and more balanced across the evaluation dimensions, highlighting an important direction for auditory-specific editing techniques.



\section{Related Work}
\subsection{Knowledge Editing}

Knowledge editing~\citep{zhu2024causal} aims to efficiently modify the models' knowledge while avoiding catastrophic forgetting that may arise from retraining directly on the target knowledge. Existing methods~\citep{mend, efk, ike, rome} adopt various strategies: using a hypernetwork to predict parameter updates for incorporating new knowledge~\citep{efk, mend}, identifying and adjusting neurons associated with specific knowledge~\citep{rome, MEMIT, anyedit}, or leveraging in-context learning (ICL) to enforce updated knowledge~\citep{ike}. Beyond correcting factual knowledge, these techniques have also been applied to bias mitigation~\citep{chen2024large}, detoxification~\citep{wang-etal-2024-detoxifying}, personalization~\citep{lu2025mobiedit}, unlearning~\citep{editing_as_unlearning}, etc.

More recently, researchers have begun to investigate factual knowledge editing in large vision-language models (LVLMs)~\citep{llava, li2023blip}. For example, \citet{mmedit} introduced the MMEdit benchmark to explore editing visual factual knowledge, while subsequent works like VLKEB~\citep{vlkeb} and MC-MKE~\citep{mcmke} expanded the evaluation scope to provide a more comprehensive understanding of editing in visual modalities. However, no prior work has examined editing auditory attribute knowledge in LALMs, which involves abstract perceptual and auditory concepts rather than concrete facts, distinguishing SAKE from existing research.

\subsection{Large Audio-Language Models (LALMs)}
LALMs extend text-based LLMs to auditory modalities such as speech and audio, opening new possibilities for auditory understanding~\citep{gong_ltuas, tang2024salmonn, qwenaudio, qwen2audio, lu2025desta25Audio, af3}. These models typically integrate auditory encoders~\citep{whisper} with an LLM backbone~\citep{touvron2023llama, yang2024qwen2technicalreport} through fine-tuning. While these works advance the integration of auditory knowledge into LLMs, little attention has been given to how auditory-specific knowledge can be edited, which motivates our study.

\section{SAKE: Speech \& Audio Attribute Knowledge Editing Benchmark}
\label{headings}
\subsection{Problem Formulation}
Given an LALM $f$ with parameters $\theta$ and an editing dataset $\mathcal{D}_{edit} = \{(a_e, x_e, y_e)\}$, where $a_e$ denotes the auditory input, $x_e$ the text input, and $y_e$ the desired edit target, knowledge editing aims to update the model such that the edited parameters $\theta'$ enable the LALM to faithfully generate the edit target: $f(a_e, x_e; \theta') = y_e$.

We focus on editing auditory attribute knowledge within LALMs, including their perception and understanding of speaker gender, emotion, spoken language, and animal sounds. Here, $y_e$ corresponds to new auditory attribute labels (e.g., emotions or languages) that differ from the original attribute labels $y_o$ associated with $a_e$. For example, given a speech labeled with a happy emotion, we may edit the LALM so that it instead perceives the speech as sad.

For comprehensive evaluation, we introduce the four evaluation dimensions of SAKE and the corresponding metrics, followed by dataset construction details for each dimension.


\subsection{Evaluation Dimensions and Corresponding Metrics}
\label{sec:metrics}

We introduce the four dimensions of knowledge editing, namely reliability, generality, locality, and portability, together with their corresponding evaluation metrics.

\paragraph{Reliability.} The \textbf{reliability} metric $S_{rel}$ measures the proportion of editing instances in $\mathcal{D}_{edit}$ for which the edited model correctly generates the corresponding edit target. It reflects how consistently the editing method updates the model in the desired manner, and is defined as
\begin{equation}
S_{rel} = \mathbb{E}_{(a_e, x_e, y_e)\sim\mathcal{D}_{edit}} \Bigl[\mathbb{I}\bigl(f(a_e, x_e; \theta') = y_e\bigr)\Bigr],
\end{equation}
where $\mathbb{I}$ denotes the indicator function, which returns $1$ if the condition holds and $0$ otherwise.

\paragraph{Generality.} The edited models should not only generate the correct edit target for the editing data itself but also produce consistent outputs for equivalent neighborhoods of the editing data, such as speech samples sharing the same emotion as $a_e$ or paraphrased variants of $x_e$. This requirement is quantified by the \textbf{generality} metric $S_{gen}$:
\begin{equation}
S_{gen} = \mathbb{E}_{\substack{(a_e, x_e, y_e)\sim\mathcal{D}_{edit}\\(a_e', x_e') \sim \mathcal{N}(a_e, x_e)}} \Bigl[\mathbb{I}\bigl(f(a_e', x_e'; \theta') = y_e\bigr)\Bigr],
\end{equation}
where $\mathcal{N}(a_e, x_e)$ denotes the aforementioned equivalent neighborhood of the editing data $(a_e, x_e)$.

\paragraph{Locality.} While updating the edit target, the edit should also preserve unrelated knowledge to avoid unintended side effects. The \textbf{locality} metric $S_{loc}$ evaluates how well an editing method maintains the model’s knowledge outside the editing scope. Given a set of out-of-scope data $\mathcal{L}(a_e, x_e, y_e) = \{(a_\ell, x_\ell, y_\ell)\}$, consisting of auditory inputs, text inputs, and ground-truth labels, $S_{loc}$ is defined as the proportion of out-of-scope data where the model’s behavior remains unchanged after editing:
\begin{equation}
\begin{aligned}
S_{loc} =
\mathbb{E}_{\substack{
(a_e, x_e, y_e)\sim\mathcal{D}_{edit}\\
(a_\ell, x_\ell, y_\ell)\sim\mathcal{L}(a_e, x_e, y_e)
}}
\Bigl[
\mathbb{I}\bigl(
f(a_\ell, x_\ell; \theta') \\= f(a_\ell, x_\ell; \theta)
\bigr)
\Bigr].
\end{aligned}
\vspace{-3mm}
\end{equation}

Note that the locality metric evaluates whether the post-edit model preserves the knowledge and behavior on data irrelevant to the edit, rather than the accuracy on out-of-scope instances. For locality with respect to purely textual abilities, we set $a_\ell = \text{None}$, as no auditory input is involved.


\paragraph{Portability.} Knowledge is not completely disentangled or isolated but rather interconnected. Editing one piece of knowledge may influence other related one. For example, if we edit an LALM’s perception of a frog’s sound to that of a dog, the model’s knowledge of the corresponding physical characteristics of that animal should also be updated. The \textbf{portability} metric $S_{port}$ evaluates how well the editing generalizes the updated knowledge to other related knowledge:
\begin{equation}
S_{port} = \mathbb{E}_{\substack{(a_e, x_e, y_e)\sim\mathcal{D}_{edit}\\(a_p, x_p, y_p)\sim\mathcal{P}(a_e, x_e, y_e)}}
\Bigl[\mathbb{I}\bigl(f(a_p, x_p; \theta') = y_p\bigr)\Bigr].
\end{equation}

$\mathcal{P}(a_e, x_e, y_e)$ denotes the set of data connected to the edited knowledge. $a_p$, $x_p$, and $y_p$ represent the auditory input, text input, and ground-truth labels of connected instances.

All the alignment and consistency in the definition of these metrics are determined with LLM-as-a-judge~\cite{chiang-lee-2023-large}, which is detailed in Sec.~\ref{sec:evaluator} and Appendix~\ref{sec:appx_prompts}.

\subsection{Dataset Construction}

We introduce the \textbf{SAKE} benchmark to evaluate the knowledge editing methods on editing the auditory attribute knowledge in LALMs with respect to the metrics detailed in Sec.~\ref{sec:metrics}. We benchmark the editing methods on LALMs with speech and audio multiple-choice question answering. 

\paragraph{Auditory Attributes and Audio Sources.} SAKE focuses on various auditory attributes: speaker gender, speaker emotion, spoken language, and animal sound. These attributes are fundamental to speech and audio understanding, widely used in real-world applications, and consistently evaluated in LALM benchmarks~\citep{huang2025dynamicsuperb, sakshi2025mmau, sakura}. We source audios and attribute labels from SAKURA benchmark~\citep{sakura}.

\paragraph{Editing Pairs and Reliability Dataset.}
For each attribute, we construct editing pairs $(y_o, y_e)$ by uniformly sampling distinct original and target labels, resulting in 300 balanced editing pairs per attribute. To avoid bias, we ensure that all labels for a given attribute appear with approximately equal frequency as both $y_o$ and $y_e$. Each pair defines an editing instance $(a_e, x_e, y_e)$, where $a_e$ is an audio sample labeled with $y_o$ and $x_e$ is a corresponding attribute recognition question from SAKURA. In total, we construct 1,200 editing instances, forming the editing dataset $\mathcal{D}_{edit}$, which is used both to apply edits and to evaluate reliability.



\paragraph{Generality Dataset Construction.} For each editing instance $(a_e, x_e, y_e)$ in $\mathcal{D}_{edit}$, we construct its equivalent neighborhood $\mathcal{N}(a_e, x_e)$ by sampling an alternative audio $a_e'$ with the same attribute label as $a_e$ from the dataset and by paraphrasing the text question $x_e$ into $x_e'$. The paraphrased data are manually verified. Based on these variations, we create testing instances for evaluating generality, considering three cases: (1) \textbf{Type 1}: Textual equivalent neighborhood $(a_e, x_e')$; (2) \textbf{Type 2}: Auditory equivalent neighborhood $(a_e', x_e)$; (3) \textbf{Type 3}: Equivalent neighborhood involving both auditory and text modalities $(a_e', x_e')$. By incorporating these types of testing data, we comprehensively assess how well the editing methods extend the edited knowledge across the equivalent neighborhood.

\paragraph{Locality Dataset Construction.} For each editing instance $(a_e, x_e, y_e)$ in $\mathcal{D}_{edit}$, we build an out-of-scope set $\mathcal{L}(a_e, x_e, y_e)={(a_\ell, x_\ell, y_\ell)}$.

When $a_\ell \neq \text{None}$, we consider four types of auditory knowledge locality (Audio Locality): (1) \textbf{Type 1}: Editing one attribute should not affect others. We sample SAKURA question–answer (QA) pairs requiring recognition of attributes different from the edited one;
(2) \textbf{Type 2}: Within the same attribute, knowledge regarding labels not involved in the edit (neither $y_o$ nor $y_e$) should remain unchanged. We sample QA pairs targeting the same attribute with ground truth different from both $y_o$ and $y_e$, except for gender due to its binary labels;
(3) \textbf{Type 3}: When editing from $y_o$ to $y_e$, the model’s original knowledge of $y_e$ should be preserved. We evaluate this using QA pairs with the same attribute and ground truth $y_e$;
(4) \textbf{Type 4}: Editing should not disrupt general auditory capability. We draw pairs from Dynamic-SUPERB Phase-2~\citep{huang2025dynamicsuperb}, excluding tasks involving the edited attributes to ensure irrelevance to the edited knowledge.

When $a_e=\text{None}$, we evaluate text locality with textual QA pairs from MMLU~\citep{mmlu}.

\paragraph{Portability Dataset Construction.} For each instance $(a_e, x_e, y_e)$ in $\mathcal{D}_{edit}$, we construct a set of connected knowledge $\mathcal{P}(a_e, x_e, y_e)$ associated with the edited attribute. As SAKURA is designed to evaluate the integration of world knowledge with auditory attributes, it provides attribute-linked knowledge (e.g., physical characteristics of animal labels). Building on this, we design questions that probe $\mathcal{P}(a_e, x_e, y_e)$. To ensure unambiguous evaluation, we exclude questions whose answers are valid for both $y_o$ and $y_e$ (e.g., avoiding ``tail" when editing from dog to cat), ensuring that portability genuinely requires knowledge updating. 

\paragraph{Training Dataset Construction and Dataset Summary.} We construct a training dataset for editing methods that require auxiliary training data or access to data beyond test instances. It follows the same procedures as the reliability, generality, and locality datasets, with training and testing sets strictly disjoint to prevent leakage. Statistics are summarized in Table~\ref{tab:dataset-split-summary}, with additional details in Appendix~\ref{sec:appx_statistics}.

\begin{table}
    \centering
    \footnotesize
    \caption{Summary statistics of SAKE dataset splits.}
    \label{tab:dataset-split-summary}
    \resizebox{\linewidth}{!}{
        \begin{tabular}{lccc}
            \toprule
            Split & \# Instances & \# Speech/Audio Inputs & \# QA Pairs \\
            \midrule
            Train & 4{,}000 & 31{,}000 & 35{,}000 \\
            Test  & 1{,}200 & 10{,}500 & 11{,}700 \\
            \bottomrule
        \end{tabular}
    }
    \vspace{-4mm}
\end{table}

\section{Experimental Settings}
\label{others}
We use three LALMs, DeSTA2.5-Audio~\citep{lu2025desta25Audio}, Qwen2-Audio-Instruct~\citep{qwen2audio}, and Audio Flamingo 3~\cite{af3}, chosen for their strong benchmark performance~\citep{huang2025dynamicsuperb, sakura, sakshi2025mmau, DBLP:journals/corr/abs-2410-17196, lu2025speechifeval}. We refer to them as DeSTA, Qwen, and AF, respectively. Details about the model choice are in Appendix~\ref{sec:design_choice}. We apply greedy decoding and assess editing methods under two settings: \textit{single editing} and \textit{sequential editing}.

\subsection{Editing Methods}

We evaluate eight knowledge editing methods that are widely adopted in prior works~\citep{mmedit, vlkeb, mcmke, mmkebench, chen2025attribution, ma2025comprehendedit}, focusing on their effectiveness in modifying abstract auditory attribute knowledge. These methods span diverse methodological paradigms, including fine-tuning-based, hypernetwork-driven, optimization-centric, memory-augmented, and in-context editing approaches. This coverage enables systematic evaluation across a wide range of editing methods. We provide an overview of these methods below, with implementation details deferred to Appendix~\ref{sec:appx_implementation}.

\textbf{Fine-tuning} is a common approach for adapting pre-trained models to new knowledge. Following prior work on LVLMs~\citep{mmedit, vlkeb, mcmke}, we compare fine-tuning two parts in LALMs: the last layer of the LLM backbone (\textbf{FT (LLM)}) and the modality connector between the audio encoder and the LLM backbone (\textbf{FT (Audio)}). \textbf{Knowledge Editor (KE)}~\citep{efk} and \textbf{MEND}~\citep{mend} trains a hypernetwork to transform the fine-tuning gradients into parameter updates for the edit. \textbf{UnKE}~\citep{unke} optimizes specific neurons of the chosen layers to produce the edit target. \textbf{In-Context Knowledge Editing (IKE)}~\citep{ike} leverages in-context learning (ICL) to enforce knowledge updates. We consider two variants: \textbf{Instruction-based IKE (I-IKE)}, which encodes edits solely through natural language instructions in system prompts, and \textbf{Instruction+Example IKE (IE-IKE)}, which provides auditory examples retrieved from the training set. \textbf{WISE}~\citep{wise} edits models using a dual-memory design that keeps pretrained FFN weights intact while storing edit-specific updates in a lightweight side memory.

\subsection{Single Editing and Sequential Editing}

\textbf{Single editing} evaluates the performance of editing a single piece of knowledge; \textbf{sequential editing} evaluates the performance after applying several edits continuously on different knowledge, which better reflects real-world scenarios. 

For sequential editing, we construct ten independent sequences, each with ten editing instances sampled from the data used for single editing. An editing sequence is denoted as $\mathcal{S} = \{ (a_e^{(t)}, x_e^{(t)}, y_o^{(t)}, y_e^{(t)}) \}_{t=1}^{10}$, where $(a_e^{(t)}, x_e^{(t)}, y_e^{(t)})$ is sampled from $\mathcal{D}_{edit}$ and the original label $y_o^{(t)}$ is retrieved from our audio dataset. Here, $t$ indexes the order of edits within the sequence.



\begin{figure}
    \centering
    \includegraphics[width=1.0\linewidth]{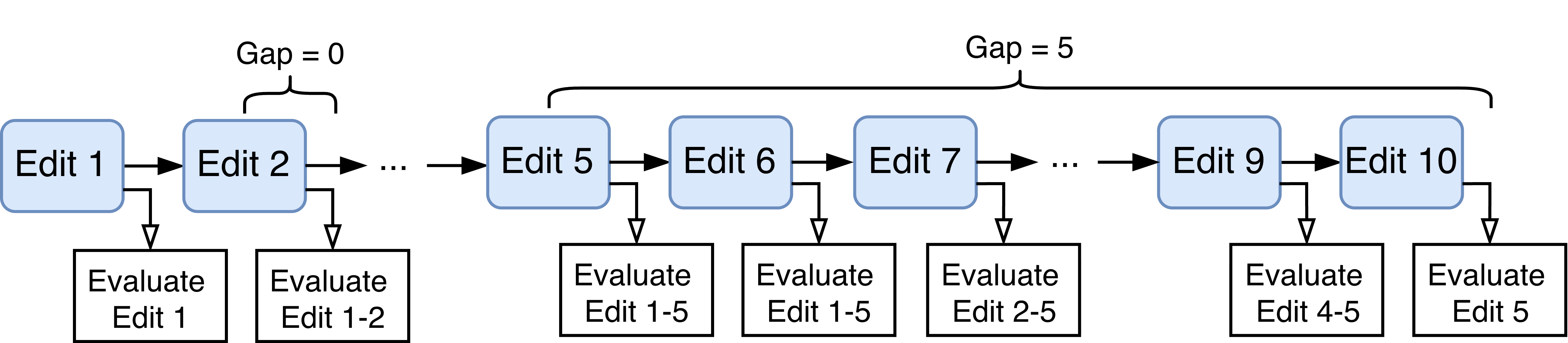}
    \caption{Example of sequential editing. For comparability, only the first five edits are evaluated, with the evaluation gap capped at five. For instance, Edit 2 is evaluated with a gap of 0, while after Edit 10, only Edit 5 is evaluated with a gap of 5.}
    \label{fig:sequential-editing-illustration}
\end{figure}

To measure how long an edit remains effective under subsequent edits, we define the \emph{gap} as the number of editing steps between when an edit is applied and when it is evaluated (Figure~\ref{fig:sequential-editing-illustration}). If an edit is introduced at step $j$ and evaluated at step $i$, then the gap is $i - j$. For consistency, we only consider the first five edits and require the gap to be at most five. This restriction keeps the number of samples comparable across gaps, since larger gaps naturally yield fewer available evaluations. To guarantee the validity of edit sequences, we impose two rules when sampling: (1) \textbf{Editing pair independence:} All original and edited labels in the sequence are mutually distinct, i.e., $y_o^{(1)}, y_e^{(1)}, \dots, y_o^{(10)}, y_e^{(10)}$ appear only once within the sequence. This avoids contradictions that could compromise the evaluation of edits with subsequent ones (e.g., editing “dog sounds” to “cat sounds” and later editing “dog sounds” to “frog sounds”); and (2) \textbf{Audio locality independence:} For each sequential editing instance $(a_e^{(t)}, x_e^{(t)}, y_o^{(t)}, y_e^{(t)})$, all samples in its audio locality dataset $\mathcal{L}(a_e^{(t)}, x_e^{(t)}, y_e^{(t)})$, denoted by $(a_\ell^{(t)}, x_\ell^{(t)}, y_\ell^{(t)})$ where $a_\ell^{(t)}\neq \text{None}$, are unrelated to the original labels of all edited instances $y_o^{(1..10)}$, thereby ensuring independent evaluation of the current edit.

\subsection{Evaluator}
\label{sec:evaluator}


Because LALMs often generate descriptive responses, we adopt LLM-as-a-judge~\citep{chiang-lee-2023-large} to compute the metrics introduced in Sec.~\ref{sec:metrics}, using GPT-5 mini as the evaluator. To validate these judgments, we conduct human evaluation on 420 randomly sampled cases, achieving 98.10\% agreement, indicating strong robustness. Additional details and prompts are provided in Appendix~\ref{sec:appx_prompts}.

\input{tables/single_main}
\section{Results}
\subsection{Single Editing}
\label{sec:single_results}


\begin{figure}[ht!]
    \centering
    \begin{subfigure}{\linewidth}
        \centering
        \includegraphics[width=\linewidth]{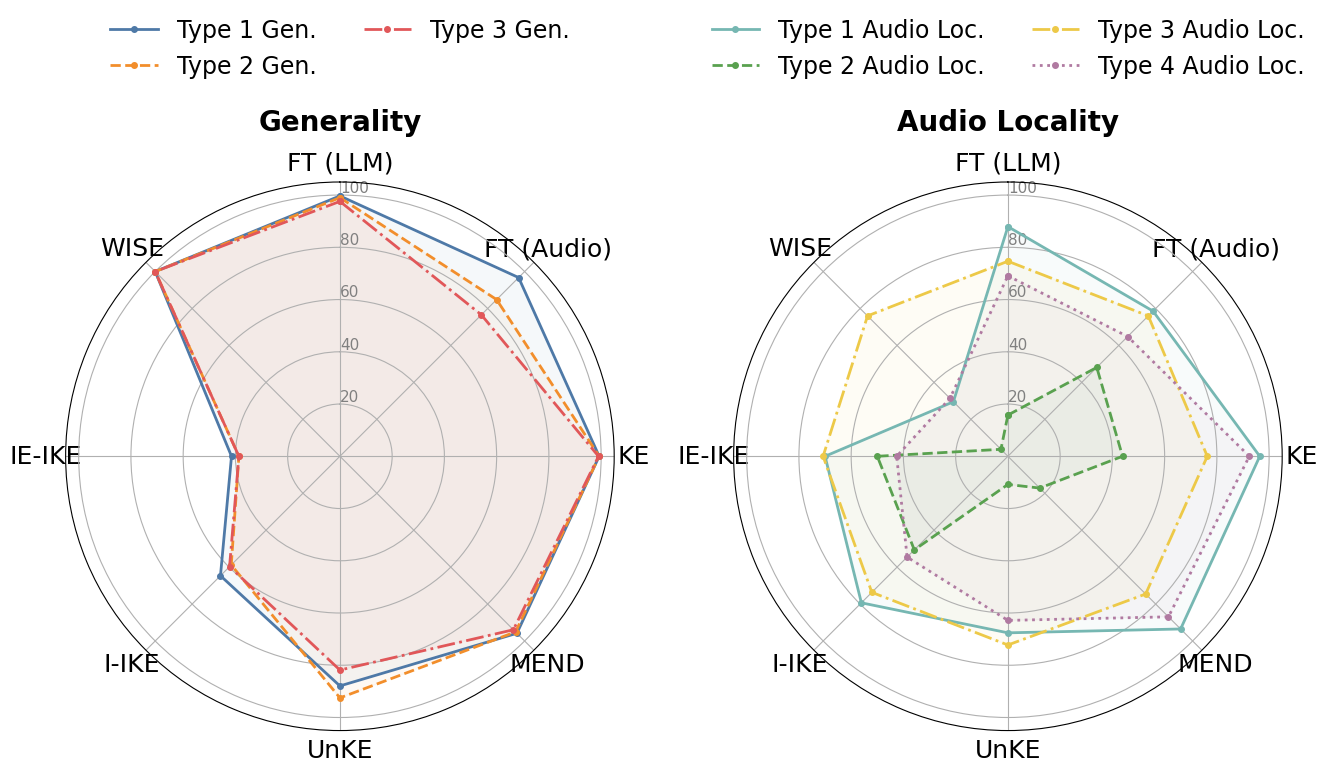}
        \caption{Results for DeSTA.}
        \label{fig:desta-breakdown}
    \end{subfigure}

    \vspace{1em} 

    \begin{subfigure}{\linewidth}
        \centering
        \includegraphics[width=\linewidth]{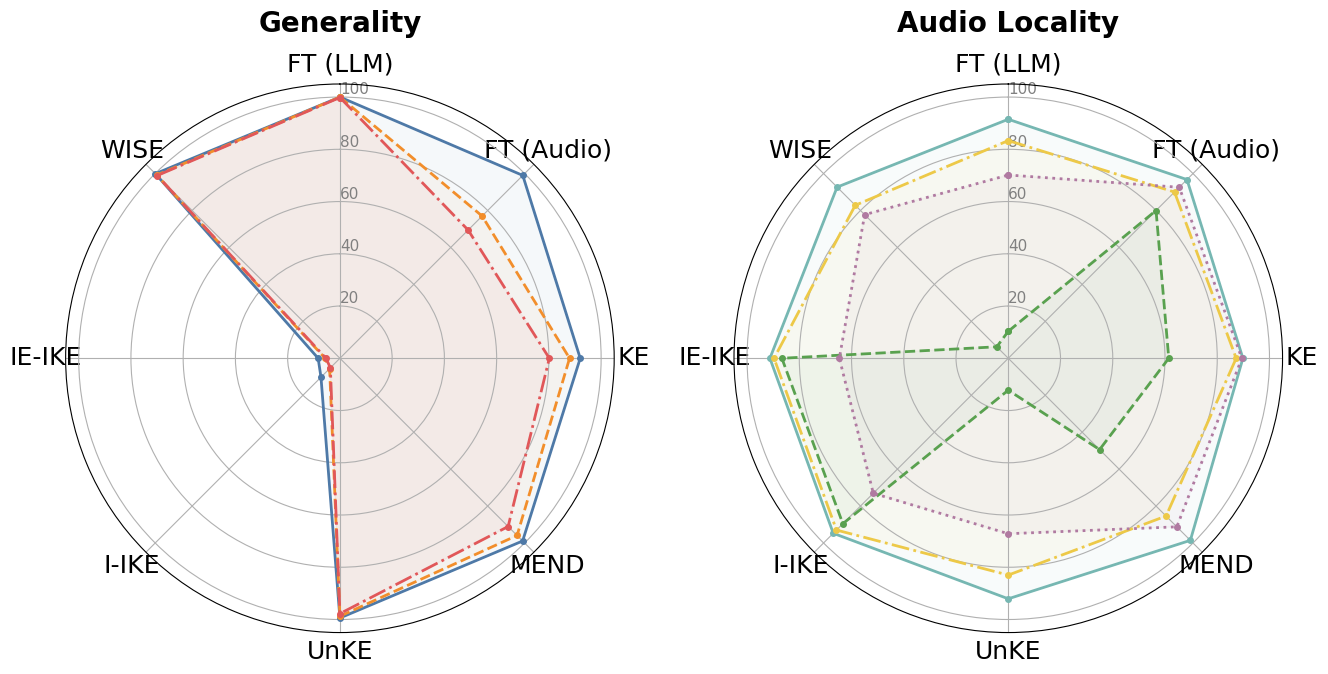}
        \caption{Results for Qwen.}
        \label{fig:qwen-breakdown}
    \end{subfigure}

    \vspace{1em} 

    \begin{subfigure}{\linewidth}
        \centering
        \includegraphics[width=\linewidth]{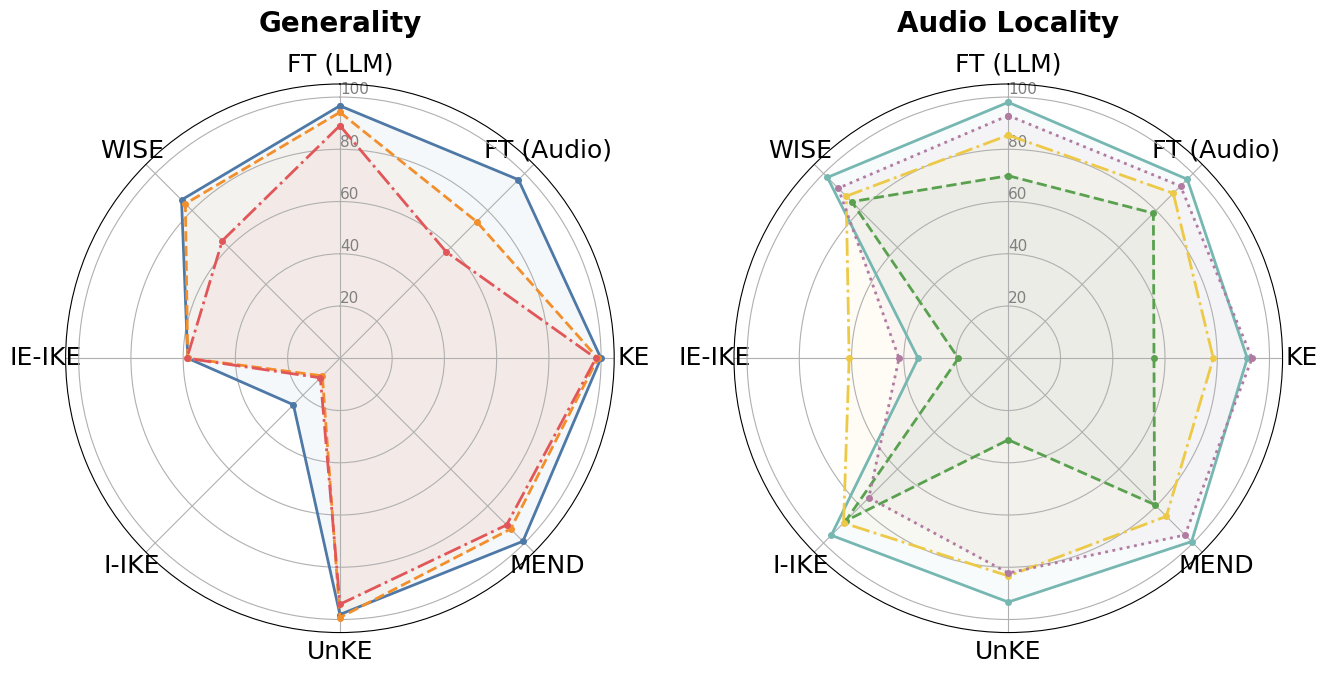}
        \caption{Results for AF.}
        \label{fig:af3-breakdown}
    \end{subfigure}

    \caption{Performances of editing methods on types of generality (left) and audio locality (right). Curves represent evaluation types.} 
    \label{fig:breakdown-generality-audio-locality}

    \vspace{-5mm}
\end{figure}

The main results of different editing methods on SAKE are shown in Table~\ref{tab:single_main}. Detailed results for each auditory attribute and relevant discussions are provided in Appendix~\ref{sec:appx_detailed_single}.





\textbf{Reliability: Parameter-updating methods are almost perfectly reliable, while IKE variants underperform.}
For all LALMs, parameter-updating methods (i.e., methods other than IKE variants) all achieve high reliability (mostly $\ge$95\% and often $\approx$100\%), showing their effectiveness in enforcing the edits. Among them, WISE consistently attains the best performance. In contrast, I-IKE and IE-IKE perform poorly. 
Interestingly, I-IKE outperforms IE-IKE on DeSTA and Qwen, despite the latter using additional auditory examples. We attribute this to those LALMs’ limited in-context learning ability: they struggle to handle \emph{multi-audio} inputs and leverage examples, unlike in LLMs~\citep{ike} and LVLMs~\citep{mmedit, vlkeb} where such methods are effective.



\textbf{Generality: Enforcing edits does not guarantee consistent generalization beyond the edited instances.}
Overall generality scores are consistently lower than reliability across all LALMs, indicating that successful edits on target instances do not readily extend to the equivalent neighborhoods. WISE and KE achieve the strongest average generality on DeSTA and AF, respectively, while FT (LLM) performs best on Qwen. Interestingly, FT (LLM) consistently outperforms FT (Audio) in average generality despite comparable reliability, suggesting that updating the LLM backbone yields better generalization than tuning the modality connector alone. We analyze the performances across types of generality in Sec.~\ref{sec:types_of_gen_loc}.

\textbf{Audio Locality: Preserving unrelated auditory knowledge remains challenging for most editing methods.}
No method consistently dominates across LALMs. On average, audio locality scores of the methods are lower than reliability and generality, indicating that enforcing auditory edits often interferes with unrelated auditory knowledge. FT (LLM) shows lower audio locality than FT (Audio), suggesting that editing different model components induces different levels of collateral effects in the auditory domain. Together with the generality results, these observations point to a trade-off between generalizing edited knowledge and preserving unrelated auditory knowledge.  
Although IKE variants do not achieve high reliability or generality, they preserve audio locality to a certain extent, likely because limited in-context learning capability constrains the propagation of the edits, albeit at the cost of weaker editing effectiveness. We discuss different types of audio locality in Sec.~\ref{sec:types_of_gen_loc}.

\textbf{Text locality: Largely preserved when edits avoid modifying the LLM backbone.}
FT (Audio) maintains perfect text locality, as it only updates the modality connector and leaves text-only behavior unaffected. In contrast, FT (LLM) substantially degrades text locality on most LALMs due to direct backbone updates. KE and MEND preserve text locality reasonably well through regularization. Other methods show larger drops, indicating that safeguarding textual capabilities requires explicit considerations during editing.

\begin{table}[ht!]
\centering
\footnotesize
\renewcommand{\arraystretch}{0.7}
\setlength{\tabcolsep}{10pt}
\caption{Standard deviation (Std.) and range of portability scores across knowledge categories for the top-two methods per model.}
\label{tab:portability_std_breakdown}
\begin{tabular}{llcc}
\toprule
\textbf{Model} & \textbf{Method} & Std. (\%) & Range (\%) \\
\midrule
\multirow{2}{*}{DeSTA} 
 & I-IKE & 12.66 & 45.13 \\
 & FT (Audio) & 22.64 & 88.18 \\
\midrule
\multirow{2}{*}{Qwen} 
 & FT (Audio) & 17.09 & 62.27 \\
 & I-IKE & 15.45 & 72.37 \\
\midrule
\multirow{2}{*}{AF} 
 & FT (Audio) & 16.64 & 65.46 \\
 & IE-IKE & 14.51 & 54.40 \\
\bottomrule
\end{tabular}%
\end{table}

\textbf{Portability: Propagating auditory edits to connected knowledge remains an open challenge.}
Overall, current editing methods do not guarantee portability when modifying auditory attribute knowledge. While no approach consistently excels, FT (Audio) exhibits comparatively more balanced portability behavior across settings. This observation suggests that edits applied through the modality connector can interact differently with related knowledge than backbone updates, though portability remains far from resolved. More broadly, the limited portability observed across methods highlights a practical challenge for scenarios where modifying an auditory attribute is expected to induce coherent updates to a broader set of related knowledge.

We further analyze portability by categorizing the related knowledge in the portability set. 
For each LALM, we select the two best-performing editing methods and report the standard deviation and range of their portability scores across knowledge categories in Table~\ref{tab:portability_std_breakdown}, with full results provided in Appendix~\ref{sec:appendix_portability_breakdown}. 
The large variation across categories shows that portability failures are not uniform: current editing methods can successfully update certain types of related knowledge while leaving others unchanged, indicating that portability exhibits substantial variation across knowledge categories rather than an all-or-nothing outcome.

\textbf{Summary.}
While most methods can enforce the desired behavior on edited instances, they struggle to generalize to equivalent ones, propagate edits to connected knowledge, and preserve unrelated knowledge within the same attribute, particularly in the auditory domain. 
Fine-tuning remains a strong baseline, with FT (Audio) showing consistently competitive performance. 
These results underscore that editing auditory attribute knowledge remains highly challenging and calls for auditory-specific editing methods.



\begin{figure*}[ht!]
    \centering
    \begin{subfigure}{0.93\textwidth}
        \centering
        \includegraphics[scale=0.2]{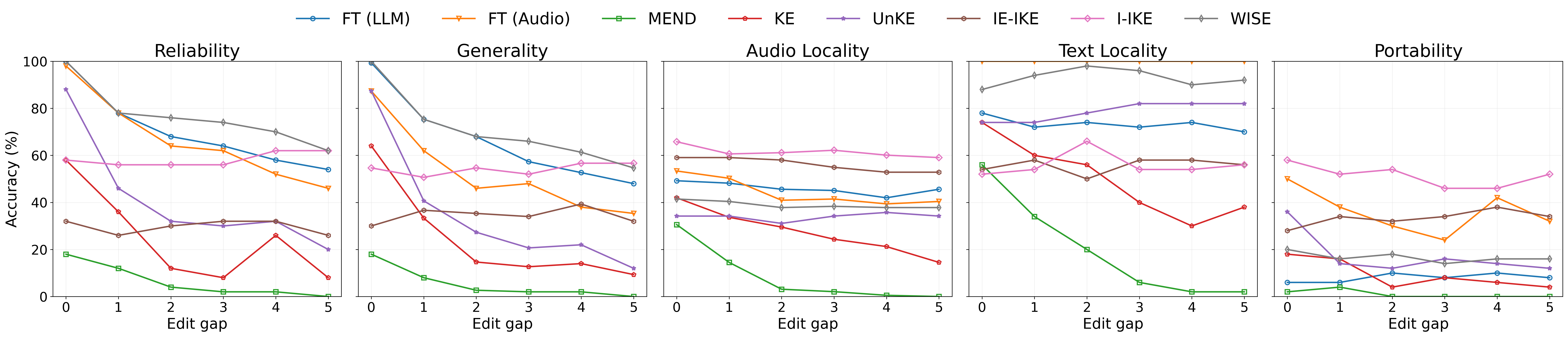}
        \vspace{-1mm}
        \caption{Results for DeSTA.}
        \label{fig:desta-sequential-editing}
    \end{subfigure}


    \begin{subfigure}{0.93\textwidth}
        \centering
        \includegraphics[scale=0.2]{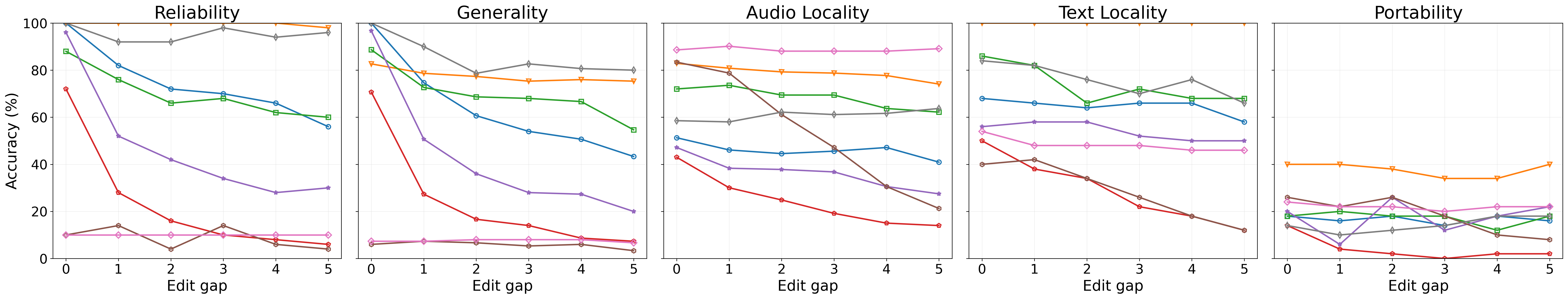}
        \vspace{-1mm}
        \caption{Results for Qwen.}
        \label{fig:qwen-sequential-editing}
    \end{subfigure}

    \begin{subfigure}{0.93\textwidth}
        \centering
        \includegraphics[scale=0.2]{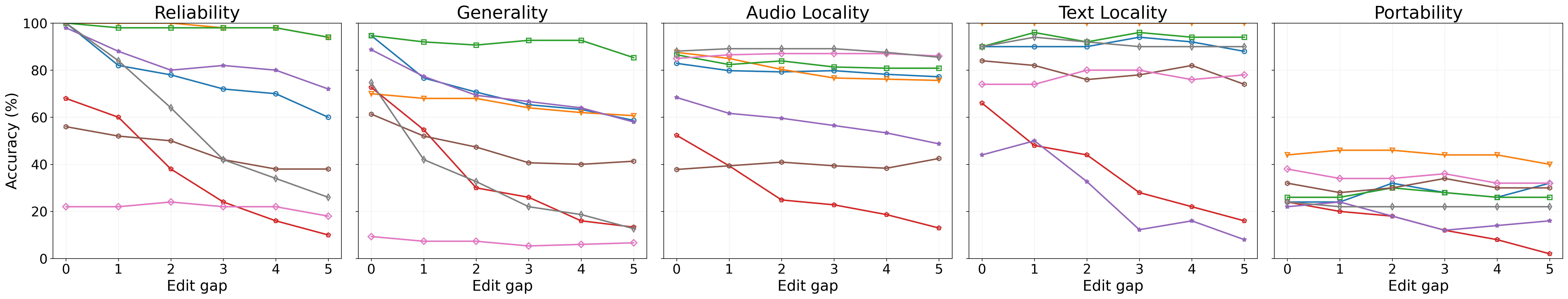}
        \vspace{-1mm}
        \caption{Results for AF.}
        \label{fig:af3-sequential-editing}
    \end{subfigure}
    \vspace{-2mm}
    \caption{Comparison of sequential editing results across models by edit gap (0–5).}
    \vspace{-4mm}
    \label{fig:sequential-editing-results}
\end{figure*}


\subsection{Detailed Analyses of Different Types of Generality and Audio Locality}
\label{sec:types_of_gen_loc}

We compare editing performance across different types of generality and audio locality in Figure~\ref{fig:breakdown-generality-audio-locality}. 
Unlike prior works in vision domains~\cite{mmedit, mcmke, vlkeb, mmkebench} that do not explicitly differentiate multiple forms of knowledge generalization and locality, our design decomposes both dimensions into distinct types, enabling more fine-grained analyses of how edits generalize and how unrelated auditory knowledge is preserved.

For generality, editing methods generally perform well on type 1 (textual neighborhood) but degrade on type 2, which requires generalization to acoustically similar yet non-identical inputs with the same labels. Type 3, which combines both textual and auditory neighborhoods, is consistently the most challenging, indicating that extending edited knowledge along the auditory dimension remains substantially more difficult than along the textual one.

For audio locality, type 2 probes knowledge within the same attribute but unrelated to the edit, and proves hardest to preserve compared to knowledge of other attributes (type 1) or the edited target itself (type 3). 
This pattern shows that modifying one aspect of an auditory attribute can inadvertently affect other aspects of the same attribute, revealing a form of intra-attribute entanglement in auditory attribute knowledge. 
Consistent with this observation, FT (LLM) exhibits substantially lower locality on type 2 than FT (Audio), despite its stronger generalization performance, suggesting a tension between generalizing edits and preserving unrelated knowledge within the same attribute. 
We provide a more detailed discussion of this phenomenon in Appendix~\ref{sec:intra_attribute}.

Type 4 assesses the preservation of general auditory processing ability. 
Most methods show clear degradation, suggesting that targeted edits often interfere with broader auditory competence. In contrast, KE and MEND better preserve original behavior, likely due to their explicit locality-oriented regularization during hypernetwork training.

\subsection{Sequential Editing}

Figure~\ref{fig:sequential-editing-results} shows the results on sequential editing.

\textbf{Significant forgetting under sequential editing in reliability and generality.} Most methods show clear declines in reliability and generality as the edit gap increases, indicating substantial forgetting of previously edited auditory knowledge. In contrast, MEND and KE degrade rapidly after only a few edits in most cases, except for MEND on AF. We attribute this behavior to their sensitivity to sequential updates, which we will discuss later.

\textbf{Sequential editing can induce degeneration.}
Some editing methods exhibit degeneration, characterized by repetitive or nonsensical outputs~\cite{degeneration, gupta-etal-2024-model, yang-etal-2024-butterfly, hsueh-etal-2024-editing}. This phenomenon is widely observed in many cases, e.g., MEND on DeSTA and Qwen, KE on all models, and UnKE on AF. These results suggest that such methods are particularly sensitive to sequential editing, where successive updates interfere with one another and induce model collapse. This highlights a key limitation of current editing approaches. Statistics and qualitative examples of degeneration are provided in Appendix~\ref{sec:case_study}. In contrast, methods like FT (Audio) show substantially less degeneration after sequential updates, showing greater robustness.

\textbf{Strength-stability trade-off in IKE variants.} Although weaker in single-edit settings, the IKE variants remain relatively stable under larger gaps, with markedly smaller degradation than parameter-updating methods; in particular, I-IKE achieves the strongest long-term reliability and generality on DeSTA. 

\textbf{No significant locality degradation under sequential editing for most methods.} Most methods do not exhibit the sharp performance drops observed in reliability and generality, indicating that sequential editing has a comparatively weaker impact on locality. Nevertheless, as discussed in Sec.~\ref{sec:types_of_gen_loc}, substantial room for improvement remains, since none of the methods consistently preserve all kinds of unrelated knowledge. Notably, KE shows a pronounced decline at larger edit gaps on all LALMs due to degeneration issues.

\textbf{Portability consistently falls below random baseline under sequential editing for most methods.} Under sequential editing, most methods demonstrate persistently poor portability, consistent with the findings reported in Sec.~\ref{sec:single_results}. In particular, most methods perform below the random baseline (27.5\%) even at small edit gaps, indicating that their ability to transfer edits to connected knowledge is limited. As a result, increasing the edit gap does not lead to substantial additional degradation or noticeable fluctuations, since portability performance is already near its lower bound.

\textbf{FT (Audio) remains a strong baseline under sequential editing.} Among all methods, FT (Audio) demonstrates relatively strong robustness across most LALMs and metrics, with notable degradation observed only in reliability and generality on DeSTA. Combined with the results in Sec.~\ref{sec:single_results}, where FT (Audio) achieves the most balanced performance in single-edit settings, these findings indicate that FT (Audio) remains a competitive baseline for auditory knowledge editing, despite its conceptual simplicity.

\section{Conclusions}

We present the first study on editing auditory attribute knowledge in LALMs and introduce SAKE, the first benchmark for evaluating this capability across four dimensions. Experiments with eight methods on three LALMs reveal challenges in preserving non-target auditory knowledge and propagating edits during reasoning. Under sequential editing, most methods suffer severe forgetting or degeneration. Overall, fine-tuning the modality connector serves as a strong and more balanced baseline for future work. In contrast, many methods that are effective in textual domains degrade substantially on auditory attributes, highlighting the unique challenges of auditory knowledge editing.

\section*{Impact Statement}

This work establishes auditory attribute knowledge editing as a previously unexplored research direction for large audio-language models (LALMs). By systematically examining the modification of perceptual auditory knowledge, we identify fundamental limitations of existing editing methods. We introduce SAKE, the first benchmark for evaluating this capability, and present a comprehensive analysis that highlights key challenges in edit reliability, knowledge generalization, preservation, and propagation. We expect SAKE to serve as a foundation for future research in this area. All related resources will be publicly released after the review process.

Knowledge editing can enhance LALMs through error correction, bias mitigation, and personalization. At the same time, modifying perceptual knowledge introduces risks of misuse, including adversarial manipulation of model behavior. This work aims to promote the responsible development and deployment of LALMs by explicitly characterizing these limitations and informing future research on principled model editing.



\bibliography{example_paper}
\bibliographystyle{icml2026}

\newpage
\appendix
\onecolumn
\section{The Usage of Large Language Models (LLMs)}
In this work, LLMs were used as judge models and as auxiliary tools for linguistic assistance, including polishing writing style, refining grammar, and proofreading. The conceptualization of the research problem, the design and construction of the benchmark, the execution of experiments, and the analysis and interpretation of results were carried out entirely by the authors without LLM involvement. All technical contributions and intellectual efforts originate from the authors, with LLMs serving only to support evaluation and to improve the clarity and readability of the manuscript.

\section{Dataset Details}
\subsection{Audio Sources}
Audios are primarily sourced from the SAKURA benchmark~\citep{sakura}, which compiles data from CommonVoice~\citep{commonvoice}, CREMA-D~\citep{cao2014crema}, ESC-50~\citep{esc50}, and Animal-Sound Dataset~\citep{animalsound}.
In addition, we supplement SAKURA with a subset of samples drawn from Dynamic-SUPERB Phase-2~\citep{huang2025dynamicsuperb} for type 4 audio locality data.

\subsection{Dataset Statistics}
\label{sec:appx_statistics}




We summarize the more details about the statistics of the training and testing sets in Tables~\ref{tab:dataset_summary_train} and~\ref{tab:dataset_summary_test}, highlighting the data diversity.

\input{tables/dataset_summary_train}
\input{tables/dataset_summary_test}

\section{Design Choices}
\label{sec:design_choice}
\subsection{Attributes}
\label{sec:appx_attribute}
SAKE covers various auditory attributes, namely animal sound, speaker emotion, speaker gender, and spoken language, because they are fundamental to speech and audio understanding, widely used in real-world applications, and consistently evaluated in existing LALM benchmarks~\citep{sakshi2025mmau, sakura, huang2025dynamicsuperb, yang-etal-2025-towards-holistic}. The attribute label sets are provided as follows:
\begin{itemize} 
    \item \textbf{Gender}: Male, Female 
    \item \textbf{Emotion}: Happy, Disgust, Sad, Fear, Angry 
    \item \textbf{Language}: English, German, Spanish, French, Italian, Chinese, Japanese, Korean 
    \item \textbf{Animal sound}: Dog, Cat, Pig, Cow, Frog, Hen, Rooster, Sheep, Crow 
\end{itemize}

As the first study on auditory knowledge editing, focusing on these core attributes provides a clear and representative starting point. Moreover, these attributes are supported by comparatively rich resources, including curated audio datasets and question–answer pairs, which are essential for constructing reliable evaluations, particularly for our portability track that involves multi-hop reasoning~\citep{sakura}. This selection establishes a focused and principled foundation for the first benchmark in this area.


\subsection{Models}


We select Qwen2-Audio~\citep{qwen2audio}, DeSTA2.5-Audio~\citep{lu2025desta25Audio}, and Audio Flamingo 3~\citep{af3} as strong, widely recognized, and representative open-source LALMs. Qwen2-Audio is widely adopted as a community baseline in studies on reasoning~\citep{sakshi2025mmau, sakura, ma2025mmar}, safety~\citep{archilles, emotional_damage}, and interpretability~\citep{yang2025audiolens}. DeSTA2.5-Audio and Audio Flamingo 3 are more recent models that achieve competitive performance across multiple benchmarks~\citep{sakshi2025mmau, huang2025dynamicsuperb, ma2025mmar, wang2025mmsu}, reflecting recent advances in LALMs.

Importantly, prior work~\citep{sakura} shows that these models already possess the world knowledge required by our portability dataset, whose connected attribute knowledge is drawn from the SAKURA benchmark. Moreover, they demonstrate state-of-the-art multi-hop reasoning that integrates auditory understanding with world knowledge. This allows observed portability errors to be attributed to \textbf{failures in edit propagation}, rather than limitations in world knowledge or inherent reasoning capability.

As our goal is to evaluate the editing methods rather than to compare LALMs themselves, we follow standard practice in knowledge editing research by applying each method to \textbf{two to three} representative models~\citep{mmedit, mmkebench, mcmke, ma2025comprehendedit, unke, wise, rome, chen2025attribution}. 
Importantly, the selected models belong to different families and differ substantially in their audio encoders, LLM backbones, modality adaptation mechanisms, and training strategies, thereby covering a diverse range of LALM designs.
The fact that our key observations consistently emerge across these models and across a wide range of editing methods suggests that the identified challenges stem from limitations of current editing techniques, rather than from idiosyncrasies of any particular model. This supports the suitability of our model selection as a testbed for evaluating auditory knowledge editing.

\section{Implementation Details of the Editing Methods}

Our implementation of editing methods is built on the EasyEdit toolkit~\citep{wang2024easyedit}.

\input{tables/hyperparams}
\input{tables/runtime.tex}

\label{sec:appx_implementation}

\subsection{Editing Methods}
\textbf{Fine-tuning} adapts the model via gradient descent on selected components. We apply it to the last layer of the LLM backbone (\textbf{FT (LLM)}) and the modality connector between audio encoders and the backbone in LALMs (\textbf{FT (Audio)}).

\textbf{Knowledge Editor (KE)}~\citep{efk} employs a hypernetwork to update parameters. It leverages a bidirectional LSTM with constrained optimization, in which generality and locality data is involved, to predict weight updates. 

\textbf{MEND}~\citep{mend} uses a hypernetwork to generate parameter updates by decomposing fine-tuning gradients into low-rank forms and transforming them into parameter updates. Similarly, it also leverages generality and locality data during training.

\textbf{UnKE}~\citep{unke} is an unstructured knowledge editing method. UnKE first finds a modified key vector by adding a small residual (delta) to the hidden state of a chosen layer so that the model’s output shifts to the desired target. In the second stage, the parameters of the chosen layer are updated to make the chosen layer naturally produce this new key vector.

\textbf{In-Context Knowledge Editing (IKE)}~\citep{ike} uses in-context learning (ICL) to modify model knowledge without parameter updates, relying on instructions and demonstrations to enforce the edited knowledge. We evaluate two variants: \textbf{Instruction-based IKE (I-IKE)} and \textbf{Instruction+Example IKE (IE-IKE)}.

\textbf{WISE}~\citep{wise} performs knowledge editing by introducing a dual-memory architecture that separates the pretrained FFN weights from a lightweight side-memory used to store edit-specific updates. During editing, WISE optimizes a routing-aware activation loss that forces edited queries to rely on the side-memory while keeping irrelevant inputs mapped to the original parameters, ensuring strong locality without degrading pretrained behavior.

\subsection{Implementation Details}
\label{sec:appx_implementation_details}

The hyperparameters of each editing method are shown in Table~\ref{tab:appx_hparams}, and their approximate execution time on NVIDIA H100 GPU is reported in Table~\ref{tab:appx_runtime}.  The execution time varies across models due to differences in their response patterns. DeSTA generally produces detailed and descriptive explanations, whereas AF typically generates concise answers without additional explanation. Qwen occasionally provides explanations, but this behavior is not consistent. 

For FT (LLM) and FT (Audio), early stopping is applied when the loss falls below 1e-2. 

For KE and MEND, which require training a hypernetwork, we train the hypernetwork for one epoch due to its rapid convergence on the validation set. 

For UnKE, the editing process consists of two stages with separate hyperparameters. The first stage uses \texttt{v\_step} iterations to find the desired key vector, with its own learning rate \texttt{v\_LR}. The second stage uses \texttt{optim\_step} iterations to update the weights of the chosen layer so that it naturally produces the new key vector, controlled by a separate learning rate \texttt{optim\_LR}. To prevent interference with unrelated knowledge, stage 2 also incorporates preservation data. Specifically, we use 20 preservation samples for text and 5 for audio. Preservation data constrains the key generator to retain original key representations while inserting new ones, preventing global parameter drift.

For WISE, the edit copy is optimized with an edit target loss and an activation-distance margin loss, where $\alpha$ constrains out-of-scope drift, $\beta$ enforces sufficient in-scope change, and $\gamma$ separates in-scope and out-of-scope activations. At inference time, an activation-based gating rule using stored edit statistics and $\mathtt{act\_ratio}$ selects between original and edited weights, ensuring edits are triggered only for in-scope inputs via preservation data.

For I-IKE and IE-IKE, the edited knowledge is encoded via natural language instructions in the system prompt, following the prompt template provided below. In IE-IKE, demonstration examples are retrieved from the training set based on the editing pair. For example, when editing the model's perception of speaker emotion from happy to sad, we retrieve an editing instance from the training set involving the same (happy, sad) editing pair. The demonstration examples illustrate how the model should respond to different types of evaluation queries, covering reliability, generality, and locality. Specifically, for each test sample, we provide 1 reliability example, 3 generality examples, 4 audio locality examples (or 3 when editing the speaker gender attribute), and 1 text locality example, all presented to the model as a multi-turn dialogue context.

\begin{tcolorbox}[title=\textbf{System Prompt Template for I-IKE and IE-IKE}, fontupper=\small]
You are performing in-context knowledge editing, but only for questions about audio inputs.\\
\\
From now on, when analyzing any audio clip, you must apply the following edits consistently:\\
\\
EDITS (New facts):\\
\verb|{edits}|\\
\\
where each entry has the form \verb|`|pre\_edit\verb|`| $\rightarrow$ \verb|`|post\_edit\verb|`|.\\
\\
Rules:\\
1. If your reasoning or prediction about an audio clip would normally lead to \verb|`|pre\_edit\verb|`|, you must instead treat it as \verb|`|post\_edit\verb|`|.\\
2. All properties, attributes, and facts that belong to \verb|`|post\_edit\verb|`| must be applied consistently, as if the audio were actually from \verb|`|post\_edit\verb|`|.\\
3. If the user’s question is unrelated to these edits, you should answer normally without making changes.\\
4. Always ensure your final answers are fully consistent with the edited mapping.\\
\end{tcolorbox}

In WISE, $\alpha$ bounds locality activation, with smaller values enforcing stricter preservation. $\beta$ sets a minimum edit activation, where larger values drive more aggressive edits. $\gamma$ enforces a margin ensuring edit
activations exceed locality activations. Finally, \texttt{act\_ratio} scales the inference routing threshold: higher values favor original weights, while lower values route more queries to the side-memory.




\section{More Details and Evaluation Prompts for LLM-as-a-judge}
\label{sec:appx_prompts}
We use GPT-5 mini (\texttt{gpt-5-mini-2025-08-07}) with minimal reasoning effort and a fixed random seed (0) as the judge in our LLM evaluator, balancing strong alignment with human judgments and cost efficiency. During evaluation, we focus on \emph{correctness} and \emph{consistency}. For reliability, generality, and portability, we assess whether the edited model's responses correctly align with the ground truth. For locality, we assess whether the edited model remains consistent with the original model’s outputs. The evaluation prompts are provided below. To further reduce costs, some cases are judged directly without invoking the LLM evaluator: empty outputs are marked as incorrect/inconsistent, and responses that exactly match the ground truth (for correctness) or the original model’s output (for consistency) are marked as correct/consistent.

Human verification was conducted on 210 randomly selected samples for correctness and consistency, respectively, yielding 98.57\% and 97.62\% agreement, indicating the robustness of the employed LLM-as-a-judge.


\newpage

\begin{tcolorbox}[title=\textbf{LLM Judge Prompt for Correctness},fontupper=\small,breakable]
\textbf{Task.} You will be given a question, a list of possible options, a ground-truth answer, and a model-generated response. Determine whether the model-generated response is \emph{correct} according to the criteria below.

\medskip
\textbf{Criteria.}
\begin{enumerate}
  \item There is one and only one correct answer. Judge \emph{incorrect} if the response does not choose any option from the list, or if it chooses more than one option.
  \item If exactly one option from the list is chosen, judge \emph{correct} if the chosen option aligns with the ground-truth answer; otherwise, judge \emph{incorrect}.
  \item Read the question, options, ground-truth answer, and model-generated response carefully before deciding.
\end{enumerate}

\medskip
\textbf{Examples.}
\begin{description}[leftmargin=1.5em]
  \item[Example 1.]
  \textit{Question:} What is the capital of France?\\
  \textit{Options:} Paris; London; Berlin; Madrid.\\
  \textit{Ground truth answer:} Paris

  \noindent\textit{Response:} ``The capital of France is Tokyo.'' $\Rightarrow$ \emph{incorrect} (does not choose any listed option).\\
  \textit{Response:} ``The capital of France is Paris and London.'' $\Rightarrow$ \emph{incorrect} (chooses more than one option).\\
  \textit{Response:} ``The capital of France is London.'' $\Rightarrow$ \emph{incorrect} (chosen option does not match ground truth).\\
  \textit{Response:} ``The capital of France is Paris.'' $\Rightarrow$ \emph{correct}.

  \item[Example 2.]
  \textit{Question:} What is the underlying emotion of the speaker?\\
  \textit{Options:} Happy; Sad; Angry; Neutral.\\
  \textit{Ground truth answer:} Happy

  \noindent\textit{Response:} ``The speaker is happy.'' $\Rightarrow$ \emph{correct}.\\
  \textit{Response:} ``The speaker expresses happiness.'' $\Rightarrow$ \emph{correct} (``happiness'' is a valid morphological variant of ``happy'').\\
  \textit{Response:} ``Happiness.'' $\Rightarrow$ \emph{correct}.
\end{description}

\medskip
\textbf{Now judge:}\\
\textit{Question:} \verb|{question}|\\
\textit{Ground truth answer:} \verb|{ground_truth}|\\
\textit{Model-generated response:} \verb|{model_response}|

\medskip
\textbf{Return format.}\\
\verb|Explanation: <Your explanation on your judgement>|\\
\verb|Judgement: <correct or incorrect>|
\end{tcolorbox}

\begin{tcolorbox}[title=\textbf{LLM Judge Prompt for Consistency}, fontupper=\small]
\textbf{Role.} You are an expert evaluator tasked with determining whether two model outputs are \emph{semantically consistent} for the same input.

\medskip
\textbf{Inputs.}\\
\textit{Question:} \verb|{question}|\\
\textit{Original Output:} \verb|{original_output}|\\
\textit{New Output:} \verb|{new_output}|

\medskip
\textbf{Task.} Decide if these two outputs are semantically consistent.

\begin{itemize}
  \item Outputs are \emph{consistent} if they express the same main message, even with different wording, phrasing, or level of detail.
  \item Outputs are \emph{inconsistent} if they contradict each other, reach different conclusions, or provide fundamentally different answers.
\end{itemize}

\medskip
\textbf{Return format.}\\
\verb|Explanation: <Your explanation on your judgement>|\\
\verb|Judgement: <consistent or inconsistent>|
\end{tcolorbox}

\FloatBarrier
\section{Detailed Results for Each Auditory Attribute under Single Editing}
\label{sec:appx_detailed_single}

\input{tables/desta_detailed_result}
\input{tables/qwen_detailed_result}
\input{tables/af3_detailed_result}

We compare the performance for each auditory attribute (\textbf{animal} sound, speaker \textbf{emotion}, speaker \textbf{gender}, and spoken \textbf{language}) under single editing here, as shown in Table~\ref{tab:desta25-main}, Table~\ref{tab:qwen2-main}, and Table~\ref{tab:af3-main} for DeSTA, Qwen, and AF, respectively.

\textbf{Reliability remains high across all attributes.} All methods except I-IKE and IE-IKE exhibit consistently high reliability across all attributes, suggesting that they can effectively update the edited knowledge on these LALMs. In contrast, the performance of I-IKE and IE-IKE varies by attributes and models, which we attribute to differences in the models’ in-context learning ability.

\textbf{Editing the modality connector limits emotional generalization.} For most methods, the generality remains relatively stable across different auditory attributes. However, FT (Audio) exhibits a notable exception: while it consistently achieves high reliability, its generality on emotion attributes degrades the most compared to its reliability on DeSTA (99.67\% $\rightarrow$ 78.56\%) and Qwen (100.00\% $\rightarrow$ 71.67\%), and the second most on AF (100.00\% $\rightarrow$ 63.00\%), among all attributes. This suggests that editing the modality connector makes it harder for the model to extend edited knowledge to similar inputs within emotion.

\textbf{Audio locality varies by attributes and models.} Most methods achieve relatively higher average performance on the gender attribute, owing to the inapplicability of type 2 audio locality, which was identified in Sec.~\ref{sec:types_of_gen_loc} as the most difficult to preserve, thereby inflating the overall average. For attributes other than gender, most methods perform comparably across attributes on Qwen. In contrast, on DeSTA, higher audio locality is observed for language in types 2 and 3, while on AF, lower audio locality is observed for emotion in type 2 across most methods. These results indicate attribute-dependent differences in preservation difficulty across LALMs. Regarding text locality, most methods demonstrate comparable performance across attributes on these LALMs.

\textbf{FT (Audio) achieves consistently high portability across attributes.} Different methods result in varying portability performance across attributes on these LALMs. A consistent observation on these models is that FT (Audio) achieves higher portability scores for all attributes, suggesting that editing the modality connector may more effectively propagate the edited knowledge to other interconnected knowledge.

\clearpage


\section{Detailed Results under Sequential Editing}

\label{sec:detailed_sequential}
We provide the statistics of the sequential editing results in Table~\ref{tab:desta_detailed_seq}, Table~\ref{tab:qwen2_detailed_seq}, and Table~\ref{tab:af3_detailed_seq}, with the corresponding line charts shown in Figure~\ref{fig:desta-sequential-editing}, Figure~\ref{fig:qwen-sequential-editing}, and Figure~\ref{fig:af3-sequential-editing} for DeSTA, Qwen, and AF, respectively.

\input{tables/desta_detailed_sequential}

\input{tables/qwen_detailed_sequential}
\input{tables/af3_detailed_seqeuntial}


\section{More Discussions of Intra-attribute Knowledge Entanglement}
\label{sec:intra_attribute}

In Sec.~\ref{sec:types_of_gen_loc}, we observe that Audio Locality Type 2 is the most difficult to preserve. We hypothesize that this failure is related to the fact that intra-attribute distinctions are often more tightly coupled than inter-attribute ones, making them more vulnerable to interference during editing. In particular, both acoustic and semantic factors may jointly contribute to this phenomenon.


\paragraph{Acoustic Entanglement} Compared to inter-attribute distinctions (e.g., emotions versus animal sounds), labels within the same attribute often share overlapping acoustic characteristics. For example, within the emotion attribute, “sad” and “fearful” speech may both exhibit lower energy, increased tension, or slower speaking rates. Such shared prosodic and spectral patterns suggest that modifying knowledge associated with one label may unintentionally affect closely related labels, making intra-attribute knowledge harder to preserve during editing.


\paragraph{Semantic Entanglement}
As shown in Figure~\ref{fig:breakdown-generality-audio-locality}, FT (LLM) performs substantially worse than FT (Audio) on Audio Locality Type 2. This gap suggests that intra-attribute distinctions are more susceptible to interference when updates are applied to the LLM backbone. A plausible explanation is that these distinctions may also rely on fine-grained semantic cues that are closely distributed within the models. Directly modifying LLM parameters may therefore increase the risk of conflating related labels.

In summary, current editing methods do not explicitly model the fine-grained structure of intra-attribute knowledge. Our results suggest that both acoustic overlap and semantic coupling may contribute to the observed failures in Audio Locality Type 2. Notably, identifying intra-attribute knowledge entanglement as a key challenge is itself an important insight enabled by our comprehensive benchmark design. Developing editing approaches that can better isolate and preserve intra-attribute distinctions remains an important direction for future work.


\clearpage

\section{Detailed Breakdown of Portability Evaluation}
\label{sec:appendix_portability_breakdown}

Given that the scope of potential related knowledge is vast and cannot be entirely enumerated, SAKE addresses this by aggregating portability evaluations across a diverse set of editing instances. Specifically, within our dataset, multiple editing instances may share the same editing pair $(y_o, y_e)$; however, they are assessed using different portability questions that target a wide variety of related concepts of the edited attribute. To provide a granular analysis of knowledge portability across these concepts, we detail the specific categories covered for each auditory attribute and present the performance breakdown under single editing.

\subsection{Taxonomy of Knowledge Categories}




\noindent
\begin{minipage}[t]{0.55\linewidth} 
    We categorize portability questions according to the underlying reasoning concepts they involve. The resulting distribution of knowledge categories is illustrated in Figure~\ref{fig:port_pie}.

    \paragraph{Animal Sounds}
    \begin{itemize}
        \item \textbf{Behavior:} Typical behavior of the animal (e.g., purring, chewing cud).
        \item \textbf{Care Item:} Required items for caring for the animal (e.g., litter box, leash).
        \item \textbf{Diet:} Dietary classification of the animal.
        \item \textbf{Family:} The closest animal in terms of biological taxonomy.
        \item \textbf{Locomotion:} The way the animal moves.
        \item \textbf{Physical:} Physical traits of the animal (e.g., claws, hooves).
        \item \textbf{Reproduction:} The reproductive method of the animal.
        \item \textbf{Vocalization:} The characteristic vocalization of the animal.
    \end{itemize}
\end{minipage}%
\hfill 
\begin{minipage}[t]{0.42\linewidth} 
    \vspace{0pt} 
    \centering
    \includegraphics[width=1.0\linewidth]{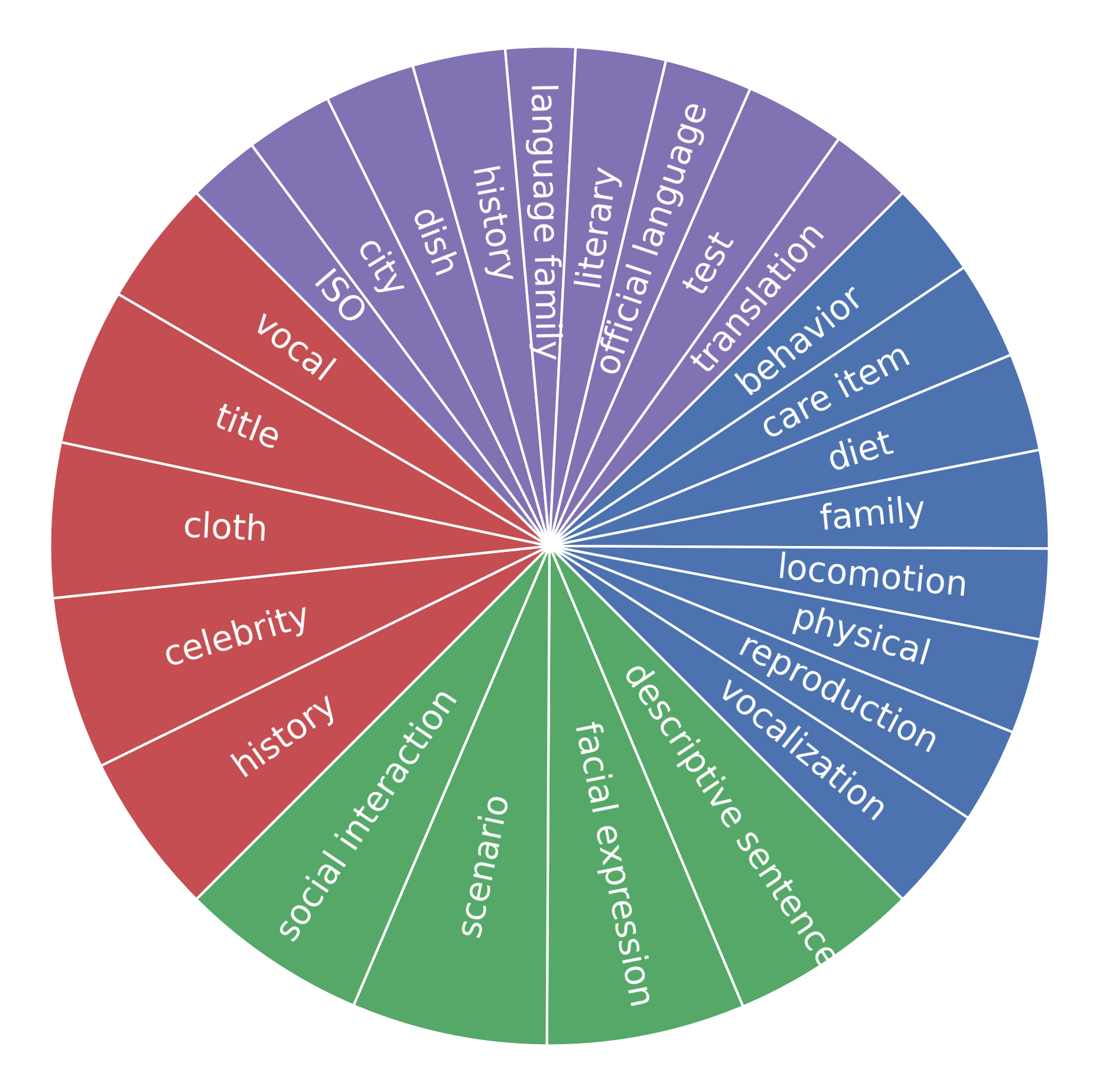}
    \captionof{figure}{Distribution of portability knowledge categories associated with edited auditory attributes.} 
    \label{fig:port_pie}
\end{minipage}

\paragraph{Speaker Emotion}
\begin{itemize}
    \item \textbf{Descriptive Sentence:} A sentence reflecting the speaker’s emotional state.
    \item \textbf{Facial Expression:} The facial expression representing the emotion.
    \item \textbf{Scenario:} The scenario or situation that matches the speaker’s emotion.
    \item \textbf{Social Interaction:} The appropriate social interaction as a response based on the emotion.
\end{itemize}

\paragraph{Speaker Gender}
\begin{itemize}
    \item \textbf{History:} A historical figure who shares the same gender as the speaker.
    \item \textbf{Celebrity:} A celebrity who shares the same gender as the speaker.
    \item \textbf{Cloth:} Traditional clothing associated with the speaker’s gender in the relevant cultural context.
    \item \textbf{Title:} Formal title of the speaker (e.g., Mr., Ms.).
    \item \textbf{Vocal:} Vocal ranges that align with the speaker’s gender.
\end{itemize}

\paragraph{Spoken Language}
\begin{itemize}
    \item \textbf{ISO:} The ISO language code of the spoken language.
    \item \textbf{City:} A city in a country where the spoken language is recognized as official.
    \item \textbf{Dish:} A dish originating from a country where the language is official.
    \item \textbf{History:} A historical figure from a country where the language is official.
    \item \textbf{Language Family:} The language family to which the language belongs.
    \item \textbf{Literary:} Books or literary works originally authored in the language.
    \item \textbf{Official Language:} The country that recognizes the spoken language as official.
    \item \textbf{Test:} A test used to evaluate proficiency in the language.
    \item \textbf{Translation:} The translation of an English word into the spoken language.
\end{itemize}

\subsection{Performance Breakdown}

Tables~\ref{tab:port_animal}, \ref{tab:port_emotion}, \ref{tab:port_gender}, and \ref{tab:port_language} detail the portability breakdown under single editting for \textit{Animal Sounds}, \textit{Speaker Emotion}, \textit{Speaker Gender}, and \textit{Spoken Language} attributes, respectively. The analysis reveals a mixed state of reasoning propagation. Rather than a binary outcome where all related concepts succeed or fail simultaneously, performance varies by category. For instance, within the \textit{Animal Sound} attribute, FT (Audio) shows a disparity in knowledge propagation, achieving 45.95\% accuracy on ``reproduction'' compared to only 7.89\% on ``family.'' Overall, instances where related concepts fail to update still constitute the majority, indicating that current editing methods struggle to consistently propagate knowledge across different reasoning dimensions.

We further emphasize that the LALMs used in this study already possess the world knowledge required by our portability dataset~\cite{sakura}, whose connected attribute knowledge is drawn from the SAKURA benchmark. Therefore, observed portability failures can be attributed to ineffective edit propagation to relevant knowledge, rather than a lack of underlying world knowledge.

\input{tables/portability_analysis/animal_sound}
\input{tables/portability_analysis/speaker_emotion}
\input{tables/portability_analysis/speaker_gender}
\input{tables/portability_analysis/spoken_language}

\clearpage

\section{Case Study}
\label{sec:case_study}

While knowledge editing in LALMs offers a way to update auditory knowledge without full retraining, the editing process is not always stable. To better understand the behaviors, we conduct a case study comparing two editing scenarios: (1) a single targeted edit applied to change an emotional attribute, and (2) sequential edits applied across multiple concepts. The comparison highlights both the strengths and limitations of current editing methods, emphasizing the trade-off between reliability of isolated edits and the accumulation of errors when multiple edits interact.

\begin{figure*}[ht]
    \centering
    \includegraphics[width=1.0\linewidth]{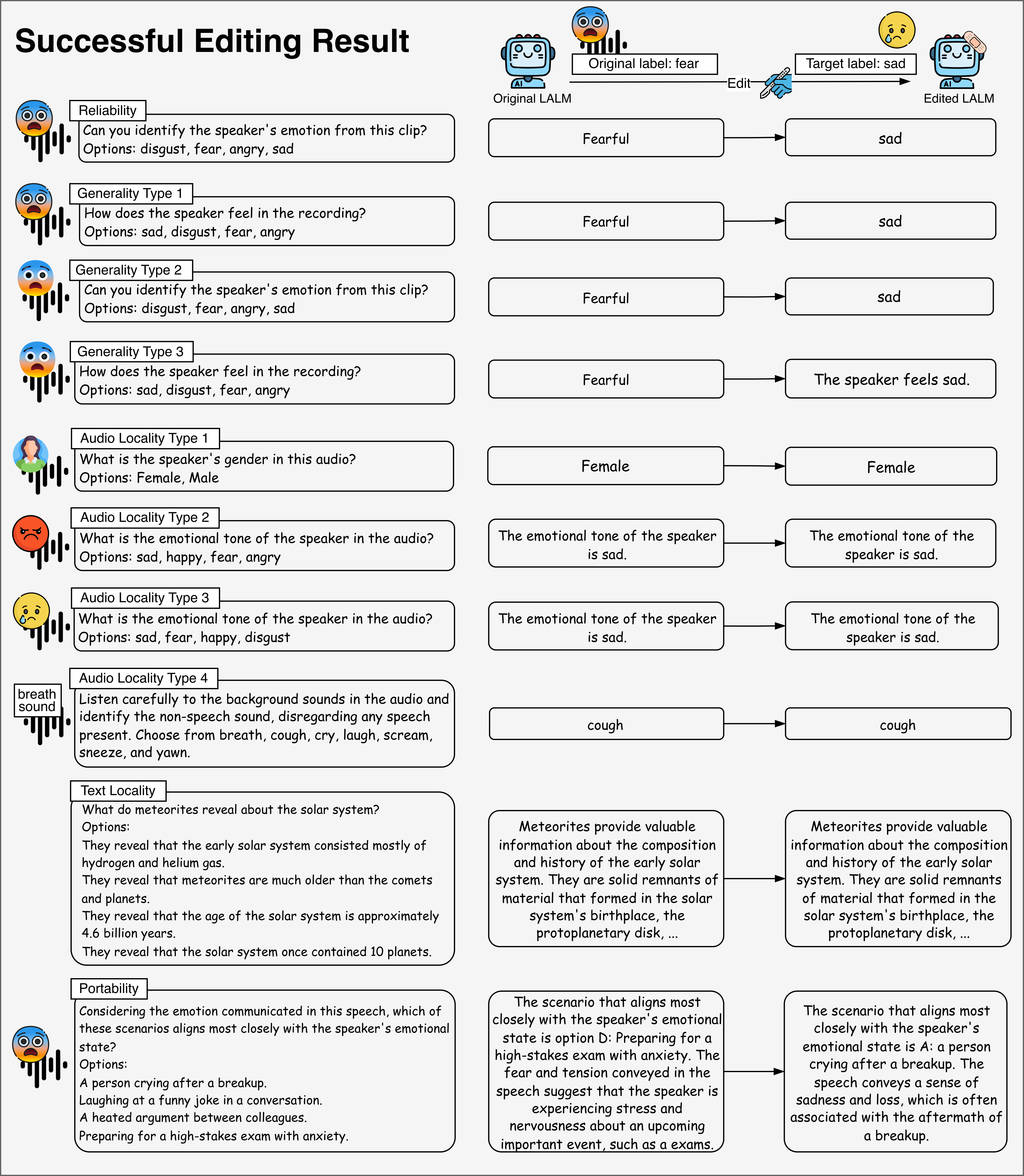}
    \caption{An example of a successful editing result by FT (Audio) on Qwen. 
    } 
    \label{fig:case_success}
\end{figure*}

\paragraph{Example of successful editing.} Figure~\ref{fig:case_success} shows the result of a successful editing example by FT (Audio) on Qwen, where we edit the model's perception of speaker emotion from “fearful” to “sad.” After editing, we observe that the model’s outputs for both reliability and generality are successfully updated to “sad.” At the same time, both audio locality and text locality remain intact, which shows that the original knowledge of the model is preserved. For portability, we observe that after the edit, the model can also perform reasoning, changing the prediction from “preparing for a high-stakes exam with anxiety” to “a person crying after a breakup.” This demonstrates that the interconnected knowledge is also updated during reasoning, and the edited model can apply this knowledge in new contexts.

\begin{figure*}[ht!]
    \centering
    \includegraphics[width=0.83\linewidth]{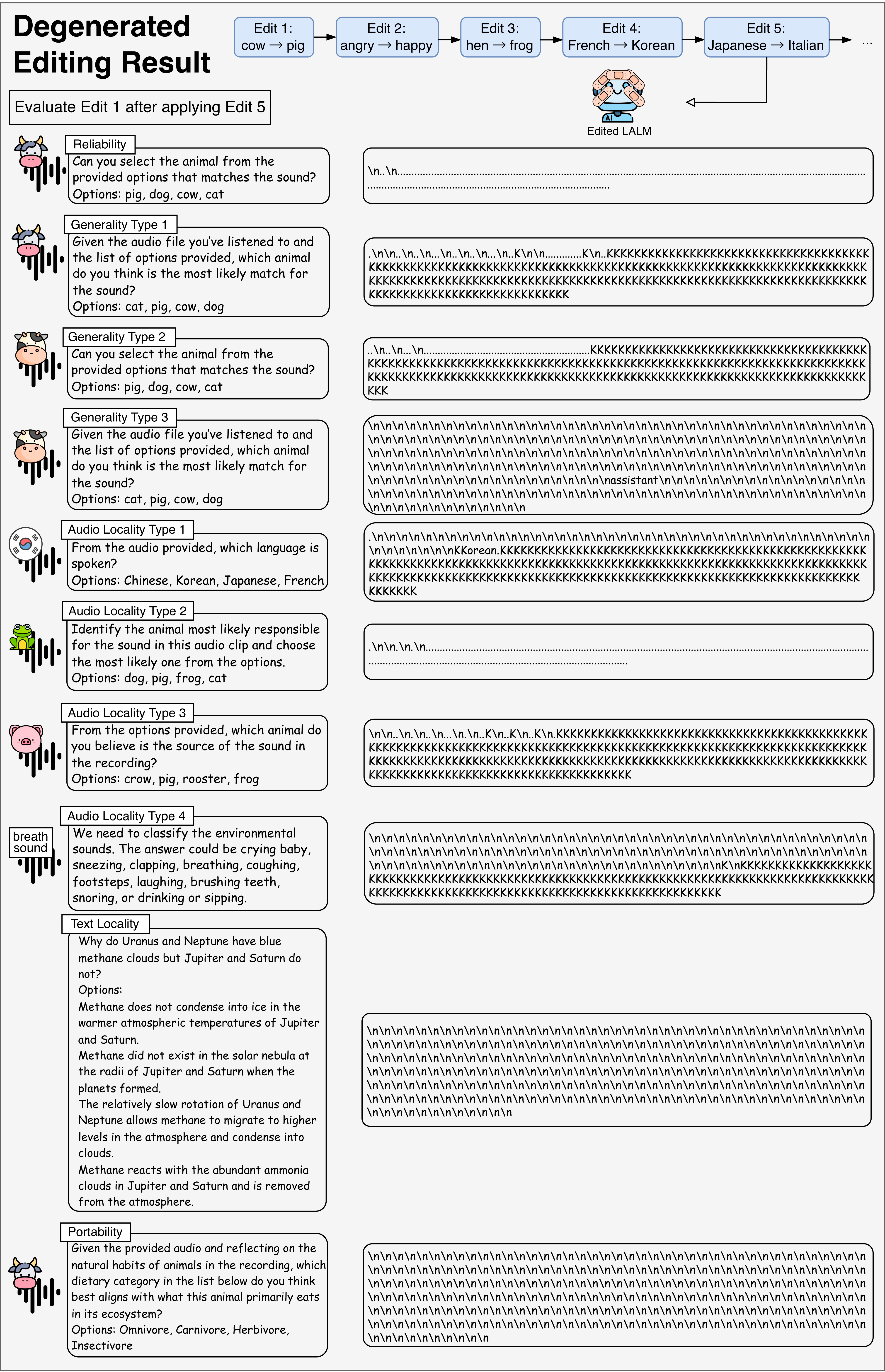}
     \caption{An example of a degenerated editing result by MEND on DeSTA. 
    }
    \label{fig:case_fail}
\end{figure*}

\paragraph{Degeneration Analysis.} Figure~\ref{fig:case_fail} shows a degenerated example from sequential editing on DeSTA with MEND. After applying multiple edits in a row, we can see that the model collapses and produces incoherent outputs such as repeated characters and newline symbols. For example, some of the output will become "happy catholic catholic catholic catholic......" or "dogdogdogdogdog...." This illustrates how sequential edits can accumulate interference and destabilize the internal representations of the model. Unlike the single-edit scenario, where the change is targeted and localized, multiple edits interact in unpredictable ways, leading to corruption of reliability, loss of generality, and failure of both locality and portability. To be practical in real-world scenarios, however, an editing method must be capable of supporting many edits simultaneously while ensuring that unrelated edits do not interfere with one another. This failure case thus underscores a key limitation of current approaches.

Quantitatively, we detect repetitive degeneration patterns in model outputs using regular expressions for each editing method and LALM under sequential editing. The frequency of degeneration is summarized in Table~\ref{tab:degeneration_ratio}. These statistics are consistent with the trends in Figure~\ref{fig:sequential-editing-results}, where MEND exhibits severe degeneration on DeSTA, KE on all models, and UnKE on AF.




\begin{table}[t]
\centering
\caption{Ratio (\%) of degenerated outputs of each editing methods on each LALM under sequential editing.}
\label{tab:degeneration_ratio}
\begin{tabular}{ccccccccc}
\toprule
\textbf{Model} & \textbf{FT (LLM)} & \textbf{FT (Audio)} & \textbf{KE} & \textbf{MEND} & \textbf{UnKE} & \textbf{WISE} & \textbf{IE-IKE} & \textbf{I-IKE} \\
\midrule
\textbf{DeSTA} & 0.30 & 1.35 & 47.77 & 44.25 & 2.23 & 0.14 & 0.00 & 0.78 \\
\textbf{Qwen}  & 0.34 & 0.20 & 11.97 & 0.85  & 2.30 & 0.20 & 1.28 & 0.34 \\
\textbf{AF}    & 0.00 & 0.00 & 9.67  & 0.00  & 6.02 & 0.00 & 0.00 & 0.00 \\
\bottomrule
\end{tabular}
\end{table}



\end{document}

%% file: tables/single_main.tex
\begin{table*}[ht]
\centering
\caption{Performance (\%) of editing methods on three models. Generality and audio locality scores are averaged across all attributes and evaluation types. Best and second-best results on individual metrics are shown in \textbf{bold} and \underline{underlined}, respectively.}
\label{tab:single_main}
\footnotesize
\setlength{\tabcolsep}{3.5pt} 
\begin{tabular}{llcccccccc}
\toprule
\textbf{Model} & \textbf{Metrics} & \textbf{FT (LLM)} & \textbf{FT (Audio)} & \textbf{KE} & \textbf{MEND} & \textbf{UnKE} & \textbf{I-IKE} & \textbf{IE-IKE} & \textbf{WISE} \\
\midrule

\multirow{5}{*}{\textbf{DeSTA}} 
& Reliability ($\uparrow$) & \underline{99.75} & 99.50 & 99.58 & 95.25 & 96.09 & 64.25 & 40.58 & \textbf{100.00} \\
& Generality ($\uparrow$) & 98.75 & 86.06 & \underline{99.17} & 94.94 & 87.42 & 61.08 & 39.61 & \textbf{100.00} \\
& Audio Locality ($\uparrow$) & 64.93 & 68.13 & \textbf{79.49} & \underline{71.44} & 57.18 & 65.49 & 58.89 & 38.90 \\
& Text Locality ($\uparrow$) & 82.67 & \textbf{100.00} & \underline{93.08} & 92.00 & 88.50 & 61.92 & 56.67 & 83.08 \\
& Portability ($\uparrow$) & 18.83 & \underline{55.33} & 17.33 & 18.08 & 17.17 & \textbf{71.33} & 34.33 & 17.34 \\
\midrule

\multirow{5}{*}{\textbf{Qwen}} 
& Reliability ($\uparrow$) & \textbf{100.00} & \textbf{100.00} & 95.33 & \textbf{100.00} & 98.67 & 10.33 & 7.83 & \textbf{100.00} \\
& Generality ($\uparrow$) & \textbf{99.94} & 81.81 & 86.69 & 95.31 & 98.64 & 7.00 & 6.44 & \underline{99.34} \\
& Audio Locality ($\uparrow$) & 67.40 & \textbf{90.49} & 83.36 & 83.29 & 68.00 & \underline{87.47} & 82.82 & 70.04 \\
& Text Locality ($\uparrow$) & 75.58 & \textbf{100.00} & 85.67 & \underline{87.25} & 72.84 & 55.67 & 51.17 & 82.09 \\
& Portability ($\uparrow$) & 24.00 & \textbf{49.92} & 26.75 & 26.25 & 26.58 & \underline{28.50} & 27.58 & 27.84 \\
\midrule

\multirow{5}{*}{\textbf{AF}} 
& Reliability ($\uparrow$) & \textbf{100.00} & \textbf{100.00} & \textbf{100.00} & 99.92 & 99.92 & 25.00 & 58.83 & \textbf{100.00} \\
& Generality ($\uparrow$) & 93.31 & 76.08 & \textbf{98.89} & \underline{93.92} & 91.17 & 15.08 & 58.36 & 77.69 \\
& Audio Locality ($\uparrow$) & 87.67 & 90.24 & 81.38 & \underline{90.64} & 76.03 & 87.07 & 40.42 & \textbf{91.18} \\
& Text Locality ($\uparrow$) & 97.00 & \textbf{100.00} & 95.83 & \underline{98.67} & 81.17 & 81.17 & 76.67 & 96.08 \\
& Portability ($\uparrow$) & 31.67 & \textbf{49.50} & 31.92 & 32.08 & 31.75 & 37.08 & \underline{40.67} & 30.75 \\
\bottomrule

\end{tabular}%
\end{table*}

%% file: tables/dataset_summary_train.tex
\begin{table}[!ht]
\centering
\caption{Dataset summary for each evaluation metric in our training dataset.}
\label{tab:dataset_summary_train}

\begin{subtable}[t]{0.49\linewidth}
\centering
\caption{Question Length (word).}
\resizebox{\linewidth}{!}{
    \begin{tabular}{cc|c|c|c|c}
        \toprule
        \multicolumn{2}{c|}{\textbf{Metric}} & \textbf{Avg.} & \textbf{Min.} & \textbf{Max.} & \textbf{Std.} \\
        \midrule    
        
        \multicolumn{2}{c|}{Reliability} & 16.40 & 12 & 22 & 2.32 \\
        \midrule
        
        \multirow{5.2}{*}{Generality} & Avg. & 16.40  & 12 & 22 & 2.32 \\
        \cmidrule{2-6}
         & Type 1 & 16.40 & 12 & 22 & 2.32 \\
         & Type 2 & 16.40 & 12 & 22 & 2.32 \\
         & Type 3 & 16.40 & 12 & 22 & 2.32 \\
        \midrule

        \multirow{5}{*}{\makecell{Audio\\Locality}} & Avg. & 26.03  & 8 & 604 & 57.53 \\
        \cmidrule{2-6}
         & Type 1 & 16.40 & 12 & 22 & 2.32 \\
         & Type 2 & 16.90 & 13 & 22 & 2.23 \\
         & Type 3 & 16.40 & 12 & 22 & 2.32 \\
         & Type 4 & 52.13 & 8 & 604 & 107.09 \\
        \midrule

        \multicolumn{2}{c|}{Text Locality} & 55.32 & 7 & 526 & 48.91 \\
        \bottomrule
    \end{tabular}
}
\end{subtable}
\hfill
\begin{subtable}[t]{0.49\linewidth}
\centering
\caption{Audio/speech duration (s).}
\resizebox{\linewidth}{!}{
    \begin{tabular}{cc|c|c|c|c}
        \toprule
        \multicolumn{2}{c|}{\textbf{Metric}} & \textbf{Avg.} & \textbf{Min.} & \textbf{Max.} & \textbf{Std.} \\
        \midrule    
        
        \multicolumn{2}{c|}{Reliability} & 4.62 & 0.13 & 17.81 & 2.27 \\
        \midrule
        
        \multirow{5.2}{*}{Generality} & Avg. & 4.62 & 0.13 & 17.81 & 2.27 \\
        \cmidrule{2-6}
         & Type 1 & 4.62 & 0.13 & 17.81 & 2.27 \\
         & Type 2 & 4.62 & 0.13 & 17.81 & 2.27 \\
         & Type 3 & 4.62 & 0.13 & 17.81 & 2.27 \\
        \midrule

        \multirow{5}{*}{\makecell{Audio\\Locality}} & Avg. & 8.27 & 0.11 & 1478.35 & 31.23 \\
        \cmidrule{2-6}
         & Type 1 & 4.62 & 0.13 & 17.81 & 2.27 \\
         & Type 2 & 4.28 & 0.13 & 17.81 & 2.22 \\
         & Type 3 & 4.62 & 0.13 & 17.81 & 2.27 \\
         & Type 4 & 18.56 & 0.11 & 1478.35 & 59.16 \\
        \bottomrule
    \end{tabular}
}
\end{subtable}

\end{table}

%% file: tables/dataset_summary_test.tex
\begin{table}[!ht]
\centering
\caption{Dataset summary for each evaluation metric in our testing dataset.}
\label{tab:dataset_summary_test}

\begin{subtable}[t]{0.49\linewidth}
\centering
\caption{Question Length (word).}
\resizebox{\textwidth}{!}{
    \begin{tabular}{cc|c|c|c|c}
        \toprule
        \multicolumn{2}{c|}{\textbf{Metric}} & \textbf{Avg.} & \textbf{Min.} & \textbf{Max.} & \textbf{Std.} \\
        \midrule    
        
        \multicolumn{2}{c|}{Reliability} & 24.23 & 7 & 48 & 10.24 \\
        \midrule
        \multirow{4}{*}{Generality} & Avg. & 23.83 & 7 & 48 & 9.99 \\
        \cmidrule{2-6}
         & Type 1 & 23.63 & 7 & 48 & 9.85 \\
         & Type 2 & 24.23 & 7 & 48 & 10.24 \\
         & Type 3 & 23.63 & 7 & 48 & 9.85 \\
        \midrule
        \multirow{5}{*}{\makecell{Audio\\Locality}} & Avg. & 31.51 & 7 & 604 & 55.81 \\
        \cmidrule{2-6}
         & Type 1 & 23.93 & 7 & 48 & 9.91 \\
         & Type 2 & 25.35 & 12 & 48 & 9.61 \\
         & Type 3 & 23.94 & 7 & 48 & 9.92 \\
         & Type 4 & 51.29 & 9 & 604 & 104.30 \\
        \midrule

        \multicolumn{2}{c|}{Text Locality} & 53.59 & 7 & 307 & 42.28 \\
        \midrule

        \multicolumn{2}{c|}{Portability} & 33.53 & 13 & 63 & 12.38 \\
        \bottomrule
    \end{tabular}
}
\end{subtable}
\hfill
\begin{subtable}[t]{0.49\linewidth}
\centering
\caption{Audio/speech duration (s).}
\resizebox{\textwidth}{!}{
    \begin{tabular}{cc|c|c|c|c}
        \toprule
        \multicolumn{2}{c|}{\textbf{Metric}} & \textbf{Avg.} & \textbf{Min.} & \textbf{Max.} & \textbf{Std.} \\
        \midrule    
        
        \multicolumn{2}{c|}{Reliability} & 4.75 & 0.24 & 20.78 & 2.43 \\
        \midrule
        
        \multirow{4}{*}{Generality} & Avg. & 5.82 & 0.24 & 20.78 & 4.98 \\
        \cmidrule{2-6}
         & Type 1 & 4.75 & 0.24 & 20.78 & 2.43 \\
         & Type 2 & 6.35 & 0.24 & 20.78 & 5.79 \\
         & Type 3 & 6.35 & 0.24 & 20.78 & 5.79 \\
         \midrule

        \multirow{5}{*}{\makecell{Audio\\Locality}} & Avg. & 8.12 & 0.17 & 759.53 & 26.98 \\
        \cmidrule{2-6}
         & Type 1 & 4.78 & 0.28 & 20.78 & 2.46 \\
         & Type 2 & 4.41 & 0.24 & 20.78 & 2.38 \\
         & Type 3 & 4.73 & 0.24 & 20.78 & 2.50 \\
         & Type 4 & 17.63 & 0.17 & 759.53 & 50.89 \\
        \midrule
        \multicolumn{2}{c|}{Portability} & 4.75 & 0.24 & 20.78 & 2.43 \\
        \bottomrule
    \end{tabular}
}
\end{subtable}

\end{table}

%% file: tables/hyperparams.tex
\begin{table}[ht!]

\caption{Hyper-parameters of each editing method. I-IKE and IE-IKE are excluded because they do not modify model parameters.}

\label{tab:appx_hparams}
\setlength{\tabcolsep}{3pt}
\centering

    \begin{tabular}{lllll}
    \toprule
    \multicolumn{5}{c}{\textbf{FT (LLM)}} \\
    \midrule
    \textbf{Model} & \textbf{Max Steps} & \textbf{Edit Layer} & \textbf{Optimizer} & \textbf{Edit LR} \\
    \midrule
    DeSTA & 15 & Layer 31 of Transformer Module & Adam & 1e-5 \\
    Qwen & 15 & Layer 31 of Transformer Module & Adam & 1e-4 \\
    AF & 15 & Layer 27 of Transformer Module & Adam & 2e-5 \\
    \midrule
    
    \multicolumn{5}{c}{\textbf{FT (Audio)}} \\
    \midrule
    \textbf{Model} & \textbf{Max Steps} & \textbf{Edit Layer} & \textbf{Optimizer} & \textbf{Edit LR} \\
    \midrule
    DeSTA & 15 & perception.connector & Adam & 1e-4 \\
    Qwen & 15 & multi\_modal\_projector & Adam & 1e-4 \\
    AF & 15 & multi\_modal\_projector & Adam & 5e-4 \\
    \midrule

    \multicolumn{5}{c}{\textbf{KE}} \\
    \midrule
    \textbf{Model} & \textbf{Epoch} & \textbf{Edit Layer} & \textbf{Optimizer} & \textbf{LR} \\
    \midrule
    DeSTA & 1 & layer 29, 30, 31 of Transformer Module & RMSprop & 3e-4 \\
    Qwen & 1 & layer 29, 30, 31 of Transformer Module & RMSprop & 3e-4 \\
    AF & 1 & layer 25, 26, 27 of Transformer Module & RMSprop & 3e-4 \\
    \midrule

    \multicolumn{5}{c}{\textbf{MEND}} \\
    \midrule
    \textbf{Model} & \textbf{Epoch} & \textbf{Edit Layer} & \textbf{Optimizer} & \textbf{LR} \\
    \midrule
    DeSTA & 1 & layer 29, 30, 31 of Transformer Module & Adam & 1e-6 \\
    Qwen & 1 & layer 29, 30, 31 of Transformer Module & Adam & 1e-6 \\
    AF & 1 & layer 25, 26, 27 of Transformer Module & Adam & 1e-6 \\
    \midrule

    \multicolumn{5}{c}{\textbf{UnKE}} \\
    \midrule
    \textbf{Model} & \textbf{\makecell{v\_step/optim\_step}} & \textbf{Edit Layer} & \textbf{preserve\_data} & \textbf{v\_LR/optim\_LR} \\
    \midrule
    DeSTA & 25/50 & layer 15 of Transformer Module & 20(text)/5(audio) & 5e-1/2e-4 \\
    Qwen    & 25/50 & layer 20 of Transformer Module & 20(text)/5(audio) & 5e-1/2e-4 \\
    AF & 25/50 & layer 20 of Transformer Module & 20(text)/5(audio) & 5e-1/2e-4 \\
    \midrule

    \multicolumn{5}{c}{\textbf{WISE}} \\
    \midrule
    \textbf{Model} & \textbf{\makecell[l]{$\alpha$, $\beta$, $\gamma$, act\_ratio}} & \textbf{Edit Layer} & \textbf{preserve\_data} & \textbf{edit\_LR/n\_iter} \\
    \midrule
    DeSTA & 2,20,10,0.88 & layer 29 of Transformer Module & 10(text)/10(audio) & 0.1/50 \\
    Qwen & 5,20,10,0.88 & layer 26 of Transformer Module & 10(text)/10(audio) & 1.0/50 \\
    AF & 5,20,10,0.00 & layer 26 of Transformer Module & 10(text)/10(audio) & 1e-2/100 \\
    \bottomrule
    \end{tabular}

\end{table}

%% file: tables/runtime.tex
\begin{table}[ht!]

\caption{Approximate execution time of each editing method on an NVIDIA H100 GPU, measured for training the trainer, single editing, and sequential editing.}
\label{tab:appx_runtime}
\centering
    \begin{tabular}{cccccc}
    \toprule
    \multirow{2.4}{*}{\textbf{Method}} & \multirow{2.4}{*}{\textbf{Model}} & \multicolumn{3}{c}{\textbf{Execution Time}} \\
    \cmidrule{3-5}
    
     & & \textbf{Training Trainer} & \textbf{Single Editing} & \textbf{Sequential Editing} \\
    \midrule
    
    \multirow{3}{*}{\textbf{FT (LLM)}}
     & DeSTA & - & 5h 50m & 2h 00m \\
     & Qwen & - & 2h 35m & 1h 15m \\
     & AF & - & 2h 00m & 0h 40m \\
    \midrule

    \multirow{3}{*}{\textbf{FT (Audio)}}
     & DeSTA & - & 4h 20m & 2h 00m \\
     & Qwen & - & 3h 00m & 1h 30m \\
     & AF & - & 1h 30m & 0h 40m \\
    \midrule

    \multirow{3}{*}{\textbf{KE}} 
     & DeSTA & 4h 10m & 8h 40m & 9h 45m \\
     & Qwen & 2h 10m & 4h 50m & 3h 20m \\
     & AF & 2h 10m & 3h 00m & 2h 15m \\
    \midrule

    \multirow{3}{*}{\textbf{MEND}}
     & DeSTA & 4h 10m & 4h 00m & 4h 10m \\
     & Qwen & 2h 20m & 2h 10m & 1h 50m \\
     & AF & 2h 20m & 2h 00m & 0h 25m \\
    \midrule

    \multirow{3}{*}{\textbf{UnKE}}
     & DeSTA & - & 3h 00m & 1h 05m\\
     & Qwen & - & 1h 35m & 0h 30m\\
     & AF & - & 1h 45m & 0h 25m\\
    \midrule

    \multirow{3}{*}{\textbf{I-IKE}}
     & DeSTA & - & 6h 00m & 3h 45m \\
     & Qwen & - & 3h 00m & 2h 00m \\
     & AF & - & 1h 20m & 0h 45m \\
    \midrule

    \multirow{3}{*}{\textbf{IE-IKE}}
     & DeSTA & - & 3h 20m & 8h 00m \\
     & Qwen & - & 2h 00m & 3h 50m \\
     & AF & - & 2h 10m & 4h 00m \\
     \midrule

     \multirow{3}{*}{\textbf{WISE}}
     & DeSTA & - & 6h 10m & 2h 00m \\
     & Qwen & - & 3h 00m & 1h 10m \\
     & AF & - & 7h 00m & 1h 45m \\
    
    \bottomrule
    \end{tabular}
\end{table}

%% file: tables/desta_detailed_result.tex

\begin{table}

\caption{Detailed results of the four metrics of each auditory attribute across different editing methods on DeSTA under single editing. Attr. denotes auditory attributes, and Port. denotes portability. For generality and audio locality, Avg. indicates the average performance across all types of the corresponding metric. (\%)}
\label{tab:desta25-main}
\setlength{\tabcolsep}{3pt}
\centering
\resizebox{\textwidth}{!}{
    \begin{tabular}{cc|c|cccc|ccccc|c|c}
    \toprule
    \multirow{2.4}{*}{\textbf{Method}} & \multirow{2.4}{*}{\textbf{Attr.}} & \multirow{2.4}{*}{\textbf{Reliability}} & \multicolumn{4}{c|}{\textbf{Generality}} & \multicolumn{5}{c|}{\textbf{Audio Locality}} & \multirow{2.4}{*}{\textbf{\makecell[c]{Text\\Locality}}} & \multirow{2.4}{*}{\textbf{Port.}} \\ [3pt]
    &  &  & Avg. & Type 1 & Type 2 & Type 3 & Avg. & Type 1 & Type 2 & Type 3 & Type 4 &  &  \\ 
    \midrule
    
    \multirow[c]{5.2}{*}{\textbf{\makecell{FT \\ (LLM)}}} & ALL & 99.75 & 98.75 & 99.67 & 99.00 & 97.58 & 64.93 & 87.92 & 15.78 & 74.67 & 69.08 & 82.67 & 18.83 \\
    \cmidrule(lr){2-14} 
     & Animal & 99.33 & 99.11 & 99.67 & 99.33 & 98.33 & 55.08 & 91.00 & 9.33 & 56.33 & 63.67 & 83.33 & 20.00 \\
     & Emotion & 100.00 & 99.56 & 99.33 & 100.00 & 99.33 & 53.17 & 80.67 & 7.33 & 59.67 & 65.00 & 77.67 & 22.00 \\
     & Gender & 99.67 & 96.78 & 99.67 & 96.67 & 94.00 & 84.44 & 90.33 & - & 88.33 & 74.67 & 88.33 & 6.00 \\
     & Language & 100.00 & 99.56 & 100.00 & 100.00 & 98.67 & 71.92 & 89.67 & 30.67 & 94.33 & 73.00 & 81.33 & 27.33 \\
    \midrule
    
    \multirow{5.2}{*}{\textbf{\makecell{FT \\ (Audio)}}} & ALL & 99.50 & 86.06 & 96.75 & 84.83 & 76.58 & 68.13 & 78.58 & 48.11 & 76.00 & 64.83 & 100.00 & 55.33 \\
    \cmidrule(lr){2-14} 
     & Animal & 99.00 & 85.67 & 97.33 & 83.00 & 76.67 & 62.08 & 81.67 & 40.67 & 60.00 & 66.00 & 100.00 & 29.33 \\
     & Emotion & 99.67 & 78.56 & 95.33 & 73.67 & 66.67 & 60.75 & 77.67 & 38.67 & 66.00 & 60.67 & 100.00 & 46.00 \\
     & Gender & 99.67 & 92.22 & 99.33 & 92.33 & 85.00 & 77.78 & 78.33 & - & 87.33 & 67.67 & 100.00 & 82.67 \\
     & Language & 99.67 & 87.78 & 95.00 & 90.33 & 78.00 & 74.33 & 76.67 & 65.00 & 90.67 & 65.00 & 100.00 & 63.33 \\
    \midrule
    
    \multirow{5.2}{*}{\textbf{KE}} & ALL & 99.58 & 99.17 & 99.25 & 99.25 & 99.00 & 79.49 & 96.42 & 43.89 & 76.33 & 92.42 & 93.08 & 17.33 \\
    \cmidrule(lr){2-14} 
     & Animal & 98.67 & 97.33 & 97.67 & 97.33 & 97.00 & 77.42 & 97.00 & 61.67 & 59.00 & 92.00 & 93.00 & 23.33 \\
     & Emotion & 100.00 & 100.00 & 100.00 & 100.00 & 100.00 & 66.25 & 98.33 & 16.33 & 59.33 & 91.00 & 93.67 & 16.00 \\
     & Gender & 99.67 & 99.56 & 99.33 & 99.67 & 99.67 & 94.22 & 96.67 & - & 92.33 & 93.67 & 93.33 & 7.67 \\
     & Language & 100.00 & 99.78 & 100.00 & 100.00 & 99.33 & 83.75 & 93.67 & 53.67 & 94.67 & 93.00 & 92.33 & 22.33 \\
    \midrule
    
    \multirow{5.2}{*}{\textbf{MEND}} & ALL & 95.25 & 94.94 & 95.83 & 95.08 & 93.92 & 71.44 & 93.50 & 17.22 & 74.58 & 86.92 & 92.00 & 18.08 \\
    \cmidrule(lr){2-14} 
     & Animal & 90.00 & 89.22 & 91.33 & 88.67 & 87.67 & 63.50 & 93.67 & 19.33 & 57.00 & 84.00 & 92.33 & 20.33 \\
     & Emotion & 99.67 & 98.44 & 99.00 & 99.00 & 97.33 & 62.17 & 94.67 & 9.00 & 59.33 & 85.67 & 93.33 & 23.67 \\
     & Gender & 98.67 & 97.00 & 97.33 & 99.33 & 94.33 & 91.78 & 97.00 & - & 88.33 & 90.00 & 90.33 & 4.67 \\
     & Language & 92.67 & 95.11 & 95.67 & 93.33 & 96.33 & 73.42 & 88.67 & 23.33 & 93.67 & 88.00 & 92.00 & 23.67 \\
    \midrule
    
    \multirow{5.2}{*}{\textbf{UnKE}} & ALL & 96.09 & 87.42 & 87.92 & 92.50 & 81.84 & 57.18 & 67.59 & 10.78 & 72.25 & 62.83 & 88.50 & 17.17 \\
    \cmidrule(lr){2-14} 
     & Animal & 99.67 & 96.11 & 98.00 & 99.67 & 90.67 & 52.75 & 69.67 & 16.33 & 60.00 & 65.00 & 88.33 & 27.33 \\
     & Emotion & 95.00 & 87.67 & 87.33 & 90.67 & 85.00 & 50.50 & 66.67 & 12.00 & 57.33 & 66.00 & 87.67 & 18.33 \\
     & Gender & 99.67 & 86.11 & 86.00 & 99.67 & 72.67 & 71.89 & 58.33 & - & 94.33 & 63.00 & 88.00 & 6.00 \\
     & Language & 90.00 & 79.78 & 80.33 & 80.00 & 79.00 & 53.58 & 75.67 & 4.00 & 77.33 & 57.33 & 90.00 & 17.00 \\
    \midrule
    
    \multirow{5.2}{*}{\textbf{I-IKE}} & ALL & 64.25 & 61.08 & 64.67 & 58.83 & 59.75 & 65.49 & 79.33 & 50.67 & 73.67 & 54.58 & 61.92 & 71.33 \\
    \cmidrule(lr){2-14} 
     & Animal & 90.33 & 92.67 & 92.67 & 92.67 & 92.67 & 54.92 & 83.67 & 22.00 & 59.67 & 54.33 & 60.67 & 78.33 \\
     & Emotion & 59.00 & 49.78 & 58.00 & 44.67 & 46.67 & 59.92 & 80.67 & 45.33 & 62.00 & 51.67 & 59.67 & 65.33 \\
     & Gender & 69.00 & 64.11 & 71.00 & 60.33 & 61.00 & 73.33 & 81.33 & - & 79.67 & 59.00 & 63.67 & 72.67 \\
     & Language & 38.67 & 37.78 & 37.00 & 37.67 & 38.67 & 75.75 & 71.67 & 84.67 & 93.33 & 53.33 & 63.67 & 69.00 \\
    \midrule
    
    \multirow{5.2}{*}{\textbf{IE-IKE}} & ALL & 40.58 & 39.61 & 41.50 & 38.67 & 38.67 & 58.89 & 70.00 & 50.00 & 70.75 & 42.58 & 56.67 & 34.33 \\
    \cmidrule(lr){2-14} 
     & Animal & 77.00 & 78.56 & 80.00 & 76.67 & 79.00 & 51.33 & 78.67 & 23.33 & 59.00 & 44.33 & 60.67 & 44.00 \\
     & Emotion & 27.00 & 26.67 & 28.67 & 25.33 & 26.00 & 51.67 & 68.00 & 33.67 & 61.33 & 43.67 & 52.33 & 28.67 \\
     & Gender & 41.67 & 37.67 & 43.00 & 36.00 & 34.00 & 61.56 & 74.33 & - & 67.67 & 42.67 & 60.67 & 32.67 \\
     & Language & 16.67 & 15.56 & 14.33 & 16.67 & 15.67 & 71.67 & 59.00 & 93.00 & 95.00 & 39.67 & 53.00 & 32.00 \\
    \midrule
    
    \multirow{5.2}{*}{\textbf{WISE}} & ALL & 100.00 & 100.00 & 100.00 & 100.00 & 100.00 & 38.90 & 29.52 & 3.78 & 76.00 & 31.38 & 83.08 & 17.34 \\
    \cmidrule(lr){2-14} 
     & Animal & 100.00 & 100.00 & 100.00 & 100.00 & 100.00 & 33.94 & 39.33 & 5.00 & 58.00 & 33.44 & 83.33 & 11.33 \\
     & Emotion & 100.00 & 100.00 & 100.00 & 100.00 & 100.00 & 24.17 & 16.33 & 6.33 & 56.33 & 17.67 & 76.00 & 2.68 \\
     & Gender & 100.00 & 100.00 & 100.00 & 100.00 & 100.00 & 63.44 & 43.67 & - & 93.67 & 53.00 & 90.00 & 29.67 \\
     & Language & 100.00 & 100.00 & 100.00 & 100.00 & 100.00 & 34.06 & 18.73 & 0.00 & 96.00 & 21.40 & 83.00 & 25.67 \\
    \bottomrule

    \end{tabular}
}

\end{table}

%% file: tables/qwen_detailed_result.tex

\begin{table}

\caption{Detailed results of the four metrics of each auditory attribute across different editing methods on Qwen under single editing. Attr. denotes auditory attributes, and Port. denotes portability. For generality and audio locality, Avg. indicates the average performance across all types of the corresponding metric. (\%)}
\label{tab:qwen2-main}
\setlength{\tabcolsep}{3pt}
\centering
\resizebox{\textwidth}{!}{
    \begin{tabular}{cc|c|cccc|ccccc|c|c}
    \toprule
    \multirow{2.4}{*}{\textbf{Method}} & \multirow{2.4}{*}{\textbf{Attr.}} & \multirow{2.4}{*}{\textbf{Reliability}} & \multicolumn{4}{c|}{\textbf{Generality}} & \multicolumn{5}{c|}{\textbf{Audio Locality}} & \multirow{2.4}{*}{\textbf{\makecell[c]{Text\\Locality}}} & \multirow{2.4}{*}{\textbf{Port.}} \\ [3pt]
    &  &  & Avg. & Type 1 & Type 2 & Type 3 & Avg. & Type 1 & Type 2 & Type 3 & Type 4 &  &  \\ 
    \midrule
    
    \multirow[c]{5.2}{*}{\textbf{\makecell{FT \\ (LLM)}}} & ALL & 100.00 & 99.94 & 100.00 & 100.00 & 99.83 & 67.40 & 91.50 & 10.33 & 83.33 & 70.17 & 75.58 & 24.00 \\
    \cmidrule(lr){2-14} 
     & Animal & 100.00 & 99.89 & 100.00 & 100.00 & 99.67 & 64.17 & 93.00 & 8.33 & 86.33 & 69.00 & 70.33 & 22.33 \\
     & Emotion & 100.00 & 100.00 & 100.00 & 100.00 & 100.00 & 58.83 & 93.33 & 8.00 & 66.00 & 68.00 & 80.33 & 19.00 \\
     & Gender & 100.00 & 100.00 & 100.00 & 100.00 & 100.00 & 83.67 & 85.33 & - & 92.33 & 73.33 & 79.67 & 26.00 \\
     & Language & 100.00 & 99.89 & 100.00 & 100.00 & 99.67 & 67.00 & 94.33 & 14.67 & 88.67 & 70.33 & 72.00 & 28.67 \\
    \midrule
    
    \multirow{5.2}{*}{\textbf{\makecell{FT \\ (Audio)}}} & ALL & 100.00 & 81.81 & 99.00 & 77.00 & 69.42 & 90.49 & 96.75 & 80.00 & 90.00 & 92.58 & 100.00 & 49.92 \\
    \cmidrule(lr){2-14} 
     & Animal & 100.00 & 84.22 & 99.00 & 80.33 & 73.33 & 94.17 & 98.00 & 93.67 & 93.33 & 91.67 & 100.00 & 48.33 \\
     & Emotion & 100.00 & 71.67 & 99.00 & 62.67 & 53.33 & 85.00 & 98.33 & 67.00 & 84.33 & 90.33 & 100.00 & 42.33 \\
     & Gender & 100.00 & 97.44 & 100.00 & 96.00 & 96.33 & 94.67 & 98.00 & - & 92.00 & 94.00 & 100.00 & 62.33 \\
     & Language & 100.00 & 73.89 & 98.00 & 69.00 & 54.67 & 89.17 & 92.67 & 79.33 & 90.33 & 94.33 & 100.00 & 46.67 \\
    \midrule
    
    \multirow{5.2}{*}{\textbf{KE}} & ALL & 95.33 & 86.69 & 92.00 & 87.92 & 80.17 & 83.36 & 89.83 & 61.44 & 87.25 & 89.42 & 85.67 & 26.75 \\
    \cmidrule(lr){2-14} 
     & Animal & 97.00 & 88.11 & 89.33 & 94.00 & 81.00 & 88.33 & 93.67 & 76.33 & 92.33 & 91.00 & 82.67 & 23.00 \\
     & Emotion & 98.33 & 76.67 & 91.33 & 74.33 & 64.33 & 77.92 & 97.33 & 52.67 & 72.00 & 89.67 & 86.67 & 22.67 \\
     & Gender & 87.00 & 89.44 & 90.33 & 89.33 & 88.67 & 84.44 & 72.67 & - & 93.67 & 87.00 & 85.67 & 38.33 \\
     & Language & 99.00 & 92.56 & 97.00 & 94.00 & 86.67 & 83.00 & 95.67 & 55.33 & 91.00 & 90.00 & 87.67 & 23.00 \\
    \midrule
    
    \multirow{5.2}{*}{\textbf{MEND}} & ALL & 100.00 & 95.31 & 98.92 & 95.92 & 91.08 & 83.29 & 98.58 & 49.44 & 85.42 & 91.25 & 87.25 & 26.25 \\
    \cmidrule(lr){2-14} 
     & Animal & 100.00 & 97.22 & 98.00 & 98.33 & 95.33 & 84.50 & 97.00 & 64.33 & 89.67 & 87.00 & 84.67 & 21.33 \\
     & Emotion & 100.00 & 89.89 & 99.00 & 92.00 & 78.67 & 75.67 & 100.00 & 42.00 & 70.33 & 90.33 & 89.67 & 23.33 \\
     & Gender & 100.00 & 95.89 & 100.00 & 94.33 & 93.33 & 94.89 & 99.00 & - & 92.33 & 93.33 & 89.33 & 36.00 \\
     & Language & 100.00 & 98.22 & 98.67 & 99.00 & 97.00 & 81.00 & 98.33 & 42.00 & 89.33 & 94.33 & 85.33 & 24.33 \\
    \midrule
    
    \multirow{5.2}{*}{\textbf{UnKE}} & ALL & 98.67 & 98.64 & 99.42 & 98.75 & 97.75 & 68.00 & 92.08 & 12.22 & 83.00 & 67.17 & 72.84 & 26.58 \\
    \cmidrule(lr){2-14} 
     & Animal & 97.00 & 97.89 & 98.67 & 95.67 & 99.33 & 63.42 & 91.00 & 9.33 & 87.00 & 66.33 & 72.67 & 20.33 \\
     & Emotion & 98.33 & 97.33 & 99.00 & 99.67 & 93.33 & 62.00 & 95.33 & 18.67 & 64.67 & 69.33 & 71.00 & 16.33 \\
     & Gender & 100.00 & 99.56 & 100.00 & 99.67 & 99.00 & 82.33 & 90.00 & - & 91.67 & 65.33 & 75.67 & 38.00 \\
     & Language & 99.33 & 99.78 & 100.00 & 100.00 & 99.33 & 64.25 & 92.00 & 8.67 & 88.67 & 67.67 & 72.00 & 31.67 \\
    \midrule
    
    \multirow{5.2}{*}{\textbf{I-IKE}} & ALL & 10.33 & 7.00 & 10.25 & 5.58 & 5.17 & 87.47 & 94.75 & 89.56 & 93.00 & 73.08 & 55.67 & 28.50 \\
    \cmidrule(lr){2-14} 
     & Animal & 7.00 & 6.89 & 6.67 & 7.33 & 6.67 & 89.75 & 96.33 & 94.33 & 94.33 & 74.00 & 57.33 & 24.00 \\
     & Emotion & 19.33 & 13.11 & 19.67 & 9.00 & 10.67 & 82.75 & 97.00 & 78.00 & 83.33 & 72.67 & 58.00 & 22.00 \\
     & Gender & 9.67 & 3.00 & 8.67 & 0.33 & 0.00 & 88.67 & 94.33 & - & 99.00 & 72.67 & 57.33 & 45.00 \\
     & Language & 5.33 & 5.00 & 6.00 & 5.67 & 3.33 & 89.00 & 91.33 & 96.33 & 95.33 & 73.00 & 50.00 & 23.00 \\
    \midrule
    
    \multirow{5.2}{*}{\textbf{IE-IKE}} & ALL & 7.83 & 6.44 & 8.33 & 5.67 & 5.33 & 82.82 & 91.25 & 86.67 & 89.67 & 64.67 & 51.17 & 27.58 \\
    \cmidrule(lr){2-14} 
     & Animal & 6.33 & 6.22 & 6.67 & 6.00 & 6.00 & 84.17 & 90.33 & 91.00 & 88.33 & 67.00 & 52.67 & 26.33 \\
     & Emotion & 9.00 & 11.56 & 11.67 & 10.67 & 12.33 & 79.50 & 96.67 & 76.33 & 80.33 & 64.67 & 54.33 & 21.67 \\
     & Gender & 11.00 & 3.78 & 10.67 & 0.67 & 0.00 & 83.44 & 89.00 & - & 99.00 & 62.33 & 52.00 & 40.00 \\
     & Language & 5.00 & 4.22 & 4.33 & 5.33 & 3.00 & 84.33 & 89.00 & 92.67 & 91.00 & 64.67 & 45.67 & 22.33 \\
    \midrule
    
    \multirow{5.2}{*}{\textbf{WISE}} & ALL & 100.00 & 99.34 & 99.92 & 99.17 & 98.92 & 70.04 & 92.67 & 6.22 & 82.84 & 77.75 & 82.09 & 27.84 \\
    \cmidrule(lr){2-14} 
     & Animal & 100.00 & 99.67 & 99.67 & 99.67 & 99.67 & 64.00 & 90.67 & 3.67 & 86.00 & 75.67 & 79.67 & 6.67 \\
     & Emotion & 100.00 & 98.78 & 100.00 & 98.67 & 97.67 & 60.58 & 93.33 & 9.67 & 65.00 & 74.33 & 82.00 & 21.67 \\
     & Gender & 100.00 & 98.89 & 100.00 & 98.33 & 98.33 & 88.89 & 94.00 & - & 91.67 & 81.00 & 83.67 & 41.00 \\
     & Language & 100.00 & 100.00 & 100.00 & 100.00 & 100.00 & 66.67 & 92.67 & 5.33 & 88.67 & 80.00 & 83.00 & 42.00 \\
    \bottomrule

    \end{tabular}
}

\end{table}

%% file: tables/af3_detailed_result.tex
\begin{table}

\caption{Detailed results of the four metrics of each auditory attribute across different editing methods on AF under single editing. Attr. denotes auditory attributes, and Port. denotes portability. For generality and audio locality, Avg. indicates the average performance across all types of the corresponding metric. (\%)}
\label{tab:af3-main}
\setlength{\tabcolsep}{3pt}
\centering
\resizebox{\textwidth}{!}{
    \begin{tabular}{cc|c|cccc|ccccc|c|c}
    \toprule
    \multirow{2.4}{*}{\textbf{Method}} & \multirow{2.4}{*}{\textbf{Attr.}} & \multirow{2.4}{*}{\textbf{Reliability}} & \multicolumn{4}{c|}{\textbf{Generality}} & \multicolumn{5}{c|}{\textbf{Audio Locality}} & \multirow{2.4}{*}{\textbf{\makecell[c]{Text\\Locality}}} & \multirow{2.4}{*}{\textbf{Port.}} \\ [3pt]
    &  &  & Avg. & Type 1 & Type 2 & Type 3 & Avg. & Type 1 & Type 2 & Type 3 & Type 4 &  &  \\ 
    \midrule
    
    \multirow[c]{5.2}{*}{\textbf{\makecell{FT \\ (LLM)}}} & ALL & 100.00 & 93.31 & 96.67 & 94.17 & 89.08 & 87.67 & 98.00 & 69.89 & 85.42 & 92.92 & 97.00 & 31.67 \\
    \cmidrule(lr){2-14} 
     & Animal & 100.00 & 91.67 & 92.33 & 97.33 & 85.33 & 89.83 & 98.33 & 79.33 & 89.67 & 92.00 & 95.67 & 29.00 \\
     & Emotion & 100.00 & 88.89 & 98.33 & 85.00 & 83.33 & 78.08 & 98.00 & 53.67 & 69.67 & 91.00 & 98.00 & 17.33 \\
     & Gender & 100.00 & 98.89 & 100.00 & 98.33 & 98.33 & 93.89 & 97.33 & - & 88.67 & 95.67 & 98.33 & 53.00 \\
     & Language & 100.00 & 93.78 & 96.00 & 96.00 & 89.33 & 90.42 & 98.33 & 76.67 & 93.67 & 93.00 & 96.00 & 27.33 \\
    \midrule
    
    \multirow{5.2}{*}{\textbf{\makecell{FT \\ (Audio)}}} & ALL & 100.00 & 76.08 & 96.67 & 74.00 & 57.58 & 90.24 & 96.92 & 78.56 & 89.33 & 93.25 & 100.00 & 49.50 \\
    \cmidrule(lr){2-14} 
     & Animal & 100.00 & 59.33 & 91.00 & 55.33 & 31.67 & 96.08 & 98.67 & 94.67 & 97.33 & 93.67 & 100.00 & 46.33 \\
     & Emotion & 100.00 & 63.00 & 99.33 & 49.33 & 40.33 & 89.25 & 98.67 & 82.33 & 83.00 & 93.00 & 100.00 & 38.00 \\
     & Gender & 100.00 & 91.44 & 99.67 & 96.67 & 78.00 & 92.89 & 96.67 & - & 87.67 & 94.33 & 100.00 & 65.67 \\
     & Language & 100.00 & 90.56 & 96.67 & 94.67 & 80.33 & 83.42 & 93.67 & 58.67 & 89.33 & 92.00 & 100.00 & 48.00 \\
    \midrule
    
    \multirow{5.2}{*}{\textbf{KE}} & ALL & 100.00 & 98.89 & 99.92 & 98.75 & 98.00 & 81.38 & 91.50 & 55.78 & 78.50 & 93.33 & 95.83 & 31.92 \\
    \cmidrule(lr){2-14} 
     & Animal & 100.00 & 99.22 & 99.67 & 99.67 & 98.33 & 85.92 & 92.00 & 70.00 & 90.00 & 91.67 & 93.33 & 28.33 \\
     & Emotion & 100.00 & 97.11 & 100.00 & 96.33 & 95.00 & 71.75 & 90.00 & 36.00 & 68.00 & 93.00 & 97.00 & 19.00 \\
     & Gender & 100.00 & 100.00 & 100.00 & 100.00 & 100.00 & 84.33 & 95.67 & - & 62.67 & 94.67 & 97.33 & 52.33 \\
     & Language & 100.00 & 99.22 & 100.00 & 99.00 & 98.67 & 84.25 & 88.33 & 61.33 & 93.33 & 94.00 & 95.67 & 28.00 \\
    \midrule
    
    \multirow{5.2}{*}{\textbf{MEND}} & ALL & 99.92 & 93.92 & 99.08 & 92.42 & 90.25 & 90.64 & 99.25 & 79.22 & 85.58 & 95.67 & 98.67 & 32.08 \\
    \cmidrule(lr){2-14} 
     & Animal & 100.00 & 97.67 & 98.33 & 99.33 & 95.33 & 91.58 & 99.67 & 81.33 & 90.00 & 95.33 & 98.67 & 30.67 \\
     & Emotion & 100.00 & 87.44 & 99.00 & 82.00 & 81.33 & 84.42 & 99.00 & 70.67 & 72.67 & 95.33 & 99.00 & 17.00 \\
     & Gender & 100.00 & 93.56 & 100.00 & 90.33 & 90.33 & 93.44 & 99.00 & - & 85.67 & 95.67 & 99.33 & 53.67 \\
     & Language & 99.67 & 97.00 & 99.00 & 98.00 & 94.00 & 93.83 & 99.33 & 85.67 & 94.00 & 96.33 & 97.67 & 27.00 \\
    \midrule
    
    \multirow{5.2}{*}{\textbf{UnKE}} & ALL & 99.92 & 91.17 & 98.09 & 99.25 & 94.17 & 76.03 & 93.33 & 31.22 & 83.33 & 82.22 & 81.17 & 31.75 \\
    \cmidrule(lr){2-14} 
     & Animal & 99.67 & 95.67 & 94.67 & 100.00 & 92.33 & 75.73 & 91.33 & 41.00 & 88.67 & 81.94 & 77.67 & 27.67 \\
     & Emotion & 100.00 & 95.33 & 99.67 & 97.00 & 89.33 & 67.00 & 93.67 & 27.67 & 67.33 & 79.33 & 80.67 & 17.33 \\
     & Gender & 100.00 & 99.67 & 100.00 & 100.00 & 99.00 & 87.21 & 94.00 & - & 84.33 & 83.28 & 85.67 & 54.00 \\
     & Language & 100.00 & 98.00 & 98.00 & 100.00 & 96.00 & 74.17 & 94.33 & 25.00 & 93.00 & 84.33 & 80.67 & 28.00 \\
    \midrule
    
    \multirow{5.2}{*}{\textbf{I-IKE}} & ALL & 25.00 & 15.08 & 25.17 & 9.50 & 10.58 & 87.07 & 95.83 & 88.00 & 89.17 & 75.50 & 81.17 & 37.08 \\
    \cmidrule(lr){2-14} 
     & Animal & 12.67 & 12.11 & 11.33 & 10.67 & 14.33 & 89.25 & 94.67 & 89.00 & 93.33 & 80.00 & 81.67 & 37.67 \\
     & Emotion & 24.00 & 9.11 & 24.67 & 1.33 & 1.33 & 83.25 & 97.00 & 83.33 & 80.00 & 72.67 & 80.67 & 22.00 \\
     & Gender & 55.33 & 32.67 & 57.33 & 20.00 & 20.67 & 86.56 & 96.00 & - & 89.00 & 74.67 & 81.00 & 59.67 \\
     & Language & 8.00 & 6.44 & 7.33 & 6.00 & 6.00 & 89.08 & 95.67 & 91.67 & 94.33 & 74.67 & 81.33 & 29.00 \\
    \midrule
    
    \multirow{5.2}{*}{\textbf{IE-IKE}} & ALL & 58.83 & 58.36 & 58.25 & 58.25 & 58.58 & 40.42 & 34.50 & 19.11 & 60.92 & 41.83 & 76.67 & 40.67 \\
    \cmidrule(lr){2-14} 
     & Animal & 38.33 & 41.11 & 42.00 & 38.33 & 43.00 & 39.67 & 37.33 & 20.00 & 53.33 & 48.00 & 78.00 & 41.00 \\
     & Emotion & 51.33 & 49.33 & 50.00 & 49.33 & 48.67 & 35.58 & 34.33 & 21.33 & 48.00 & 38.67 & 77.67 & 25.67 \\
     & Gender & 99.33 & 98.56 & 99.33 & 98.33 & 98.00 & 51.78 & 29.33 & - & 84.33 & 41.67 & 75.67 & 61.33 \\
     & Language & 46.33 & 44.44 & 41.67 & 47.00 & 44.67 & 37.50 & 37.00 & 16.00 & 58.00 & 39.00 & 75.33 & 34.67 \\
    \midrule
    
    \multirow{5.2}{*}{\textbf{WISE}} & ALL & 100.00 & 77.69 & 85.75 & 83.67 & 63.67 & 91.18 & 97.92 & 84.78 & 87.75 & 92.07 & 96.08 & 30.75 \\
    \cmidrule(lr){2-14} 
     & Animal & 100.00 & 83.00 & 80.00 & 95.67 & 73.33 & 92.99 & 97.67 & 89.67 & 90.67 & 93.98 & 97.00 & 30.00 \\
     & Emotion & 100.00 & 79.33 & 96.00 & 74.00 & 68.00 & 84.17 & 99.00 & 74.67 & 73.33 & 89.67 & 96.00 & 18.00 \\
     & Gender & 100.00 & 76.00 & 97.33 & 74.33 & 56.33 & 93.55 & 97.00 & - & 91.33 & 92.31 & 97.00 & 51.00 \\
     & Language & 100.00 & 72.44 & 69.67 & 90.67 & 57.00 & 94.00 & 98.00 & 90.00 & 95.67 & 92.33 & 94.33 & 24.00 \\
    \bottomrule

    \end{tabular}
}

\end{table}

%% file: tables/desta_detailed_sequential.tex
\begin{table}[ht]
\centering
\caption{Original result of the four metrics of different editing methods on DeSTA under sequential editing. For generality and audio locality, we present the averaged results. (\%)}
\label{tab:desta_detailed_seq}
\resizebox{0.9\textwidth}{!}{%
\setlength{\tabcolsep}{3pt}
    \begin{tabularx}{\textwidth}{Yc|Y|Y|Y|Y|Y}
    \toprule
    \textbf{Method} & \textbf{Gap} & \textbf{Reliability} & \textbf{Generality} & \textbf{Audio Locality} & \textbf{Text Locality} & \textbf{Portability} \\ \midrule
    \multirow{6}{*}{\textbf{\makecell{FT\\(LLM)}}} & 0 & 100.00 & 99.33 & 49.22 & 78.00 & 6.00 \\ 
     & 1 & 78.00 & 75.33 & 48.19 & 72.00 & 6.00 \\ 
     & 2 & 68.00 & 68.00 & 45.60 & 74.00 & 10.00 \\ 
     & 3 & 64.00 & 57.33 & 45.08 & 72.00 & 8.00 \\ 
     & 4 & 58.00 & 52.67 & 41.97 & 74.00 & 10.00 \\ 
     & 5 & 54.00 & 48.00 & 45.60 & 70.00 & 8.00 \\ \midrule
    \multirow{6}{*}{\textbf{\makecell{FT\\(Audio)}}} & 0 & 98.00 & 87.33 & 53.37 & 100.00 & 50.00 \\ 
     & 1 & 78.00 & 62.00 & 50.26 & 100.00 & 38.00 \\ 
     & 2 & 64.00 & 46.00 & 40.93 & 100.00 & 30.00 \\ 
     & 3 & 62.00 & 48.00 & 41.45 & 100.00 & 24.00 \\ 
     & 4 & 52.00 & 38.00 & 39.38 & 100.00 & 42.00 \\ 
     & 5 & 46.00 & 35.33 & 40.41 & 100.00 & 32.00 \\ \midrule
    \multirow{6}{*}{\textbf{MEND}} & 0 & 18.00 & 18.00 & 30.57 & 56.00 & 2.00 \\ 
     & 1 & 12.00 & 8.00 & 14.51 & 34.00 & 4.00 \\ 
     & 2 & 4.00 & 2.67 & 3.11 & 20.00 & 0.00 \\ 
     & 3 & 2.00 & 2.00 & 2.07 & 6.00 & 0.00 \\ 
     & 4 & 2.00 & 2.00 & 0.52 & 2.00 & 0.00 \\ 
     & 5 & 0.00 & 0.00 & 0.00 & 2.00 & 0.00 \\ \midrule
    \multirow{6}{*}{\textbf{KE}} & 0 & 58.00 & 64.00 & 41.97 & 74.00 & 18.00 \\ 
     & 1 & 36.00 & 33.33 & 33.68 & 60.00 & 16.00 \\ 
     & 2 & 12.00 & 14.67 & 29.53 & 56.00 & 4.00 \\ 
     & 3 & 8.00 & 12.67 & 24.35 & 40.00 & 8.00 \\ 
     & 4 & 26.00 & 14.00 & 21.24 & 30.00 & 6.00 \\ 
     & 5 & 8.00 & 9.33 & 14.51 & 38.00 & 4.00 \\ \midrule
    \multirow{6}{*}{\textbf{UnKE}} & 0 & 88.00 & 87.33 & 34.20 & 74.00 & 36.00 \\ 
     & 1 & 46.00 & 40.67 & 34.20 & 74.00 & 14.00 \\ 
     & 2 & 32.00 & 27.33 & 31.08 & 78.00 & 12.00 \\ 
     & 3 & 30.00 & 20.67 & 34.20 & 82.00 & 16.00 \\ 
     & 4 & 32.00 & 22.00 & 35.75 & 82.00 & 14.00 \\ 
     & 5 & 20.00 & 12.00 & 34.20 & 82.00 & 12.00 \\ \midrule
    \multirow{6}{*}{\textbf{IE-IKE}} & 0 & 32.00 & 30.00 & 59.07 & 54.00 & 28.00 \\ 
     & 1 & 26.00 & 36.67 & 59.07 & 58.00 & 34.00 \\ 
     & 2 & 30.00 & 35.33 & 58.03 & 50.00 & 32.00 \\ 
     & 3 & 32.00 & 34.00 & 54.92 & 58.00 & 34.00 \\ 
     & 4 & 32.00 & 39.33 & 52.85 & 58.00 & 38.00 \\ 
     & 5 & 26.00 & 32.00 & 52.85 & 56.00 & 34.00 \\ \midrule
    \multirow{6}{*}{\textbf{I-IKE}} & 0 & 58.00 & 54.67 & 65.80 & 52.00 & 58.00 \\ 
     & 1 & 56.00 & 50.67 & 60.62 & 54.00 & 52.00 \\ 
     & 2 & 56.00 & 54.67 & 61.14 & 66.00 & 54.00 \\ 
     & 3 & 56.00 & 52.00 & 62.18 & 54.00 & 46.00 \\ 
     & 4 & 62.00 & 56.67 & 60.10 & 54.00 & 46.00 \\ 
     & 5 & 62.00 & 56.67 & 59.07 & 56.00 & 52.00 \\ \midrule
    \multirow{6}{*}{\textbf{WISE}} & 0 & 100.00 & 100.00 & 41.45 & 88.00 & 20.00 \\ 
     & 1 & 78.00 & 75.33 & 40.41 & 94.00 & 16.00 \\ 
     & 2 & 76.00 & 68.00 & 37.82 & 98.00 & 18.00 \\ 
     & 3 & 74.00 & 66.00 & 38.34 & 96.00 & 14.00 \\ 
     & 4 & 70.00 & 61.33 & 37.82 & 90.00 & 16.00 \\ 
     & 5 & 62.00 & 54.67 & 37.82 & 92.00 & 16.00 \\ \bottomrule
    \end{tabularx}%
}
\end{table}

%% file: tables/qwen_detailed_sequential.tex
\begin{table}[ht]
\centering
\caption{Original result of the four metrics of different editing methods on Qwen under sequential editing. For generality and audio locality, we present the averaged results. (\%)}
\label{tab:qwen2_detailed_seq}
\resizebox{0.9\textwidth}{!}{%
\setlength{\tabcolsep}{3pt}
    \begin{tabularx}{\textwidth}{Yc|Y|Y|Y|Y|Y}
    \toprule
    \textbf{Method} & \textbf{Gap} & \textbf{Reliability} & \textbf{Generality} & \textbf{Audio Locality} & \textbf{Text Locality} & \textbf{Portability} \\ \midrule
    \multirow{6}{*}{\textbf{\makecell{FT\\(LLM)}}} & 0 & 100.00 & 100.00 & 51.30 & 68.00 & 18.00 \\ 
     & 1 & 82.00 & 74.67 & 46.11 & 66.00 & 16.00 \\ 
     & 2 & 72.00 & 60.67 & 44.56 & 64.00 & 18.00 \\ 
     & 3 & 70.00 & 54.00 & 45.60 & 66.00 & 14.00 \\ 
     & 4 & 66.00 & 50.67 & 47.15 & 66.00 & 18.00 \\ 
     & 5 & 56.00 & 43.33 & 40.93 & 58.00 & 16.00 \\ \midrule
    \multirow{6}{*}{\textbf{\makecell{FT\\(Audio)}}} & 0 & 100.00 & 82.67 & 82.90 & 100.00 & 40.00 \\ 
     & 1 & 100.00 & 78.67 & 80.83 & 100.00 & 40.00 \\ 
     & 2 & 100.00 & 77.33 & 79.27 & 100.00 & 38.00 \\ 
     & 3 & 100.00 & 75.33 & 78.76 & 100.00 & 34.00 \\ 
     & 4 & 100.00 & 76.00 & 77.72 & 100.00 & 34.00 \\ 
     & 5 & 98.00 & 75.33 & 74.09 & 100.00 & 40.00 \\ \midrule
    \multirow{6}{*}{\textbf{MEND}} & 0 & 88.00 & 88.67 & 72.02 & 86.00 & 18.00 \\ 
     & 1 & 76.00 & 72.67 & 73.58 & 82.00 & 20.00 \\ 
     & 2 & 66.00 & 68.67 & 69.43 & 66.00 & 18.00 \\ 
     & 3 & 68.00 & 68.00 & 69.43 & 72.00 & 18.00 \\ 
     & 4 & 62.00 & 66.67 & 63.73 & 68.00 & 12.00 \\ 
     & 5 & 60.00 & 54.67 & 62.18 & 68.00 & 18.00 \\ \midrule
    \multirow{6}{*}{\textbf{KE}} & 0 & 72.00 & 70.67 & 43.01 & 50.00 & 14.00 \\ 
     & 1 & 28.00 & 27.33 & 30.05 & 38.00 & 4.00 \\ 
     & 2 & 16.00 & 16.67 & 24.87 & 34.00 & 2.00 \\ 
     & 3 & 10.00 & 14.00 & 19.17 & 22.00 & 0.00 \\ 
     & 4 & 8.00 & 8.67 & 15.03 & 18.00 & 2.00 \\ 
     & 5 & 6.00 & 7.33 & 13.99 & 12.00 & 2.00 \\ \midrule
    \multirow{6}{*}{\textbf{UnKE}} & 0 & 96.00 & 96.67 & 47.15 & 56.00 & 20.00 \\ 
     & 1 & 52.00 & 50.67 & 38.34 & 58.00 & 6.00 \\ 
     & 2 & 42.00 & 36.00 & 37.82 & 58.00 & 26.00 \\ 
     & 3 & 34.00 & 28.00 & 36.78 & 52.00 & 12.00 \\ 
     & 4 & 28.00 & 27.33 & 30.57 & 50.00 & 18.00 \\ 
     & 5 & 30.00 & 20.00 & 27.46 & 50.00 & 22.00 \\ \midrule
    \multirow{6}{*}{\textbf{IE-IKE}} & 0 & 10.00 & 6.00 & 83.42 & 40.00 & 26.00 \\ 
     & 1 & 14.00 & 7.33 & 78.76 & 42.00 & 22.00 \\ 
     & 2 & 4.00 & 6.67 & 61.14 & 34.00 & 26.00 \\ 
     & 3 & 14.00 & 5.33 & 47.15 & 26.00 & 18.00 \\ 
     & 4 & 6.00 & 6.00 & 30.57 & 18.00 & 10.00 \\ 
     & 5 & 4.00 & 3.33 & 21.24 & 12.00 & 8.00 \\ \midrule
    \multirow{6}{*}{\textbf{I-IKE}} & 0 & 10.00 & 7.33 & 88.60 & 54.00 & 24.00 \\ 
     & 1 & 10.00 & 7.33 & 90.16 & 48.00 & 22.00 \\ 
     & 2 & 10.00 & 8.00 & 88.08 & 48.00 & 22.00 \\ 
     & 3 & 10.00 & 8.00 & 88.08 & 48.00 & 20.00 \\ 
     & 4 & 10.00 & 8.00 & 88.08 & 46.00 & 22.00 \\ 
     & 5 & 10.00 & 6.67 & 89.12 & 46.00 & 22.00 \\ \midrule
    \multirow{6}{*}{\textbf{WISE}} & 0 & 100.00 & 100.00 & 58.55 & 84.00 & 14.00 \\ 
     & 1 & 92.00 & 90.00 & 58.03 & 82.00 & 10.00 \\ 
     & 2 & 92.00 & 78.67 & 62.18 & 76.00 & 12.00 \\ 
     & 3 & 98.00 & 82.67 & 61.14 & 70.00 & 14.00 \\ 
     & 4 & 94.00 & 80.67 & 61.65 & 76.00 & 18.00 \\ 
     & 5 & 96.00 & 80.00 & 63.73 & 66.00 & 18.00 \\ \bottomrule
    \end{tabularx}%
}
\end{table}

%% file: tables/af3_detailed_seqeuntial.tex
\begin{table}[ht]
\centering
\caption{Original result of the four metrics of different editing methods on AF under sequential editing. For generality and audio locality, we present the averaged results. (\%) }
\label{tab:af3_detailed_seq}
\resizebox{0.9\textwidth}{!}{%
\setlength{\tabcolsep}{3pt}
    \begin{tabularx}{\textwidth}{Yc|Y|Y|Y|Y|Y}
    \toprule
    \textbf{Method} & \textbf{Gap} & \textbf{Reliability} & \textbf{Generality} & \textbf{Audio Locality} & \textbf{Text Locality} & \textbf{Portability} \\ \midrule
    \multirow{6}{*}{\textbf{\makecell{FT\\(LLM)}}} & 0 & 100.00 & 94.67 & 82.90 & 90.00 & 24.00 \\ 
     & 1 & 82.00 & 76.67 & 79.79 & 90.00 & 24.00 \\ 
     & 2 & 78.00 & 70.67 & 79.27 & 90.00 & 32.00 \\ 
     & 3 & 72.00 & 65.33 & 79.79 & 94.00 & 28.00 \\ 
     & 4 & 70.00 & 63.33 & 78.24 & 92.00 & 26.00 \\ 
     & 5 & 60.00 & 58.67 & 77.20 & 88.00 & 32.00 \\ \midrule
    \multirow{6}{*}{\textbf{\makecell{FT\\(Audio)}}} & 0 & 100.00 & 70.00 & 87.56 & 100.00 & 44.00 \\ 
     & 1 & 100.00 & 68.00 & 84.97 & 100.00 & 46.00 \\ 
     & 2 & 100.00 & 68.00 & 80.31 & 100.00 & 46.00 \\ 
     & 3 & 98.00 & 64.00 & 76.68 & 100.00 & 44.00 \\ 
     & 4 & 98.00 & 62.00 & 76.17 & 100.00 & 44.00 \\ 
     & 5 & 94.00 & 60.67 & 75.65 & 100.00 & 40.00 \\ \midrule
    \multirow{6}{*}{\textbf{MEND}} & 0 & 100.00 & 94.67 & 86.53 & 90.00 & 26.00 \\ 
     & 1 & 98.00 & 92.00 & 82.38 & 96.00 & 26.00 \\ 
     & 2 & 98.00 & 90.67 & 83.94 & 92.00 & 30.00 \\ 
     & 3 & 98.00 & 92.67 & 81.35 & 96.00 & 28.00 \\ 
     & 4 & 98.00 & 92.67 & 80.83 & 94.00 & 26.00 \\ 
     & 5 & 94.00 & 85.33 & 80.83 & 94.00 & 26.00 \\ \midrule
    \multirow{6}{*}{\textbf{KE}} & 0 & 68.00 & 72.67 & 52.33 & 66.00 & 24.00 \\ 
     & 1 & 60.00 & 54.67 & 39.38 & 48.00 & 20.00 \\ 
     & 2 & 38.00 & 30.00 & 24.87 & 44.00 & 18.00 \\ 
     & 3 & 24.00 & 26.00 & 22.80 & 28.00 & 12.00 \\ 
     & 4 & 16.00 & 16.00 & 18.65 & 22.00 & 8.00 \\ 
     & 5 & 10.00 & 13.33 & 12.95 & 16.00 & 2.00 \\ \midrule
    \multirow{6}{*}{\textbf{UnKE}} & 0 & 98.00 & 88.67 & 68.39 & 44.00 & 22.00 \\ 
     & 1 & 88.00 & 77.33 & 61.65 & 50.00 & 24.00 \\ 
     & 2 & 80.00 & 69.33 & 59.58 & 32.65 & 18.00 \\ 
     & 3 & 82.00 & 66.67 & 56.48 & 12.24 & 12.00 \\ 
     & 4 & 80.00 & 64.00 & 53.36 & 16.00 & 14.00 \\ 
     & 5 & 72.00 & 58.00 & 48.70 & 8.00 & 16.00 \\ \midrule
    \multirow{6}{*}{\textbf{IE-IKE}} & 0 & 56.00 & 61.33 & 37.82 & 84.00 & 32.00 \\ 
     & 1 & 52.00 & 52.00 & 39.38 & 82.00 & 28.00 \\ 
     & 2 & 50.00 & 47.33 & 40.93 & 76.00 & 30.00 \\ 
     & 3 & 42.00 & 40.67 & 39.38 & 78.00 & 34.00 \\ 
     & 4 & 38.00 & 40.00 & 38.34 & 82.00 & 30.00 \\ 
     & 5 & 38.00 & 41.33 & 42.49 & 74.00 & 30.00 \\ \midrule
    \multirow{6}{*}{\textbf{I-IKE}} & 0 & 22.00 & 9.33 & 84.97 & 74.00 & 38.00 \\ 
     & 1 & 22.00 & 7.33 & 86.53 & 74.00 & 34.00 \\ 
     & 2 & 24.00 & 7.33 & 87.05 & 80.00 & 34.00 \\ 
     & 3 & 22.00 & 5.33 & 87.05 & 80.00 & 36.00 \\ 
     & 4 & 22.00 & 6.00 & 87.05 & 76.00 & 32.00 \\ 
     & 5 & 18.00 & 6.67 & 86.01 & 78.00 & 32.00 \\ \midrule
    \multirow{6}{*}{\textbf{WISE}} & 0 & 100.00 & 74.67 & 88.08 & 90.00 & 24.00 \\ 
     & 1 & 84.00 & 42.00 & 89.12 & 94.00 & 22.00 \\ 
     & 2 & 64.00 & 32.67 & 89.12 & 92.00 & 22.00 \\ 
     & 3 & 42.00 & 22.00 & 89.12 & 90.00 & 22.00 \\ 
     & 4 & 34.00 & 18.67 & 87.56 & 90.00 & 22.00 \\ 
     & 5 & 26.00 & 12.67 & 85.49 & 90.00 & 22.00 \\ \bottomrule
    \end{tabularx}%
}
\end{table}

%% file: tables/portability_analysis/animal_sound.tex
\begin{table}
\centering
\caption{Breakdown of portability performance across different knowledge categories for the \textit{Animal Sound} attribute. (\%)}
\label{tab:port_animal}

\resizebox{\textwidth}{!}{%
\begin{tabular}{cc|cccccccc}
\toprule
\textbf{Model} & \textbf{Method} & \textbf{Behavior} & \textbf{Care Item} & \textbf{Diet} & \textbf{Family} & \textbf{Locomotion} & \textbf{Physical} & \textbf{Reproduction} & \textbf{Vocalization} \\ \midrule

\multirow[c]{11.5}{*}{\textbf{DeSTA}}

 & \textbf{\makecell{FT \\ (LLM)}} & 2.70 & 23.68 & 21.05 & 2.63 & 68.57 & 13.51 & 16.22 & 15.00 \\ \cmidrule{2-10} 
 & \textbf{\makecell{FT \\ (Audio)}} & 24.32 & 39.47 & 21.05 & 5.26 & 25.71 & 32.43 & 45.95 & 40.00 \\ \cmidrule{2-10} 
 & \textbf{KE} & 10.81 & 39.47 & 5.26 & 26.32 & 34.29 & 24.32 & 8.11 & 37.50 \\ \cmidrule{2-10} 
 & \textbf{MEND} & 5.41 & 36.84 & 39.47 & 10.53 & 34.29 & 10.81 & 18.92 & 7.50 \\ \cmidrule{2-10} 
 & \textbf{UnKE} & 21.62 & 55.26 & 26.32 & 5.26 & 28.57 & 16.22 & 18.92 & 45.00 \\ \cmidrule{2-10} 
 & \textbf{I-IKE} & 83.78 & 92.11 & 68.42 & 47.37 & 85.71 & 78.38 & 78.38 & 92.50 \\ \cmidrule{2-10} 
 & \textbf{IE-IKE} & 32.43 & 57.89 & 42.11 & 13.16 & 51.43 & 37.84 & 35.14 & 80.00 \\ \cmidrule{2-10} 
 & \textbf{WISE} & 2.70 & 15.79 & 15.79 & 0.00 & 17.14 & 10.81 & 27.03 & 2.50 \\
\midrule

\multirow[c]{11.5}{*}{\textbf{Qwen}}

 & \textbf{\makecell{FT \\ (LLM)}} & 21.62 & 26.32 & 13.16 & 23.68 & 31.43 & 21.62 & 10.81 & 30.00 \\ \cmidrule{2-10} 
 & \textbf{\makecell{FT \\ (Audio)}} & 56.76 & 71.05 & 36.84 & 28.95 & 37.14 & 51.35 & 43.24 & 60.00 \\ \cmidrule{2-10} 
 & \textbf{KE} & 24.32 & 31.58 & 13.16 & 23.68 & 31.43 & 10.81 & 8.11 & 40.00 \\ \cmidrule{2-10} 
 & \textbf{MEND} & 18.92 & 42.11 & 15.79 & 15.79 & 11.43 & 16.22 & 18.92 & 30.00 \\ \cmidrule{2-10} 
 & \textbf{UnKE} & 40.54 & 21.05 & 18.42 & 10.53 & 0.00 & 8.11 & 10.81 & 50.00 \\ \cmidrule{2-10} 
 & \textbf{I-IKE} & 29.73 & 34.21 & 21.05 & 18.42 & 31.43 & 10.81 & 5.41 & 40.00 \\ \cmidrule{2-10} 
 & \textbf{IE-IKE} & 27.03 & 50.00 & 21.05 & 21.05 & 22.86 & 18.92 & 5.41 & 42.50 \\ \cmidrule{2-10} 
 & \textbf{WISE} & 18.92 & 18.42 & 0.00 & 0.00 & 2.86 & 0.00 & 0.00 & 12.50 \\
\midrule

\multirow[c]{11.5}{*}{\textbf{AF}}

 & \textbf{\makecell{FT \\ (LLM)}} & 35.14 & 39.47 & 26.32 & 18.42 & 20.00 & 32.43 & 32.43 & 27.50 \\ \cmidrule{2-10} 
 & \textbf{\makecell{FT \\ (Audio)}} & 54.05 & 60.53 & 28.95 & 28.95 & 40.00 & 48.65 & 51.35 & 57.50 \\ \cmidrule{2-10} 
 & \textbf{KE} & 37.84 & 39.47 & 23.68 & 23.68 & 17.14 & 32.43 & 24.32 & 27.50 \\ \cmidrule{2-10} 
 & \textbf{MEND} & 35.14 & 47.37 & 21.05 & 28.95 & 22.86 & 32.43 & 29.73 & 27.50 \\ \cmidrule{2-10} 
 & \textbf{UnKE} & 35.14 & 26.32 & 23.68 & 23.68 & 17.14 & 21.62 & 35.14 & 37.50 \\ \cmidrule{2-10} 
 & \textbf{I-IKE} & 40.54 & 52.63 & 23.68 & 26.32 & 31.43 & 35.14 & 43.24 & 47.50 \\ \cmidrule{2-10} 
 & \textbf{IE-IKE} & 48.65 & 47.37 & 28.95 & 31.58 & 34.29 & 35.14 & 45.95 & 55.00 \\ \cmidrule{2-10} 
 & \textbf{WISE} & 29.73 & 47.37 & 21.05 & 26.32 & 25.71 & 29.73 & 35.14 & 25.00 \\
 \bottomrule
\end{tabular}%
}

\end{table}

%% file: tables/portability_analysis/speaker_emotion.tex
\begin{table}
\centering
\caption{Breakdown of portability performance across different knowledge categories for the \textit{Speaker Emotion} attribute. (\%)}
\label{tab:port_emotion}

\begin{tabular}{cc|cccc}
\toprule
\textbf{Model} & \textbf{Method} & \textbf{Descriptive Sentence} & \textbf{Facial Expression} & \textbf{Scenario} & \textbf{Social Interaction} \\ \midrule

\multirow[c]{11.5}{*}{\textbf{DeSTA}}

 & \textbf{\makecell{FT \\ (LLM)}} & 24.32 & 19.48 & 17.11 & 27.40 \\ \cmidrule{2-6} 
 & \textbf{\makecell{FT \\ (Audio)}} & 44.59 & 46.75 & 47.37 & 45.21 \\ \cmidrule{2-6} 
 & \textbf{KE} & 17.57 & 14.29 & 11.84 & 20.55 \\ \cmidrule{2-6} 
 & \textbf{MEND} & 22.97 & 23.38 & 21.05 & 27.40 \\ \cmidrule{2-6} 
 & \textbf{UnKE} & 16.22 & 24.68 & 13.16 & 19.18 \\ \cmidrule{2-6} 
 & \textbf{I-IKE} & 72.97 & 51.95 & 68.42 & 68.49 \\ \cmidrule{2-6} 
 & \textbf{IE-IKE} & 45.95 & 23.38 & 19.74 & 26.03 \\ \cmidrule{2-6} 
 & \textbf{WISE} & 1.35 & 1.30 & 5.26 & 2.78 \\
\midrule

\multirow[c]{11.5}{*}{\textbf{Qwen}}

 & \textbf{\makecell{FT \\ (LLM)}} & 18.92 & 16.88 & 10.53 & 30.14 \\ \cmidrule{2-6} 
 & \textbf{\makecell{FT \\ (Audio)}} & 47.30 & 44.16 & 35.53 & 42.47 \\ \cmidrule{2-6} 
 & \textbf{KE} & 21.62 & 15.58 & 22.37 & 31.51 \\ \cmidrule{2-6} 
 & \textbf{MEND} & 24.32 & 15.58 & 22.37 & 31.51 \\ \cmidrule{2-6} 
 & \textbf{UnKE} & 13.51 & 12.99 & 10.53 & 28.77 \\ \cmidrule{2-6} 
 & \textbf{I-IKE} & 22.97 & 16.88 & 15.79 & 32.88 \\ \cmidrule{2-6} 
 & \textbf{IE-IKE} & 21.62 & 18.18 & 18.42 & 28.77 \\ \cmidrule{2-6} 
 & \textbf{WISE} & 22.97 & 10.39 & 19.74 & 34.25 \\
\midrule

\multirow[c]{11.5}{*}{\textbf{AF}}

 & \textbf{\makecell{FT \\ (LLM)}} & 14.86 & 18.18 & 21.05 & 17.81 \\ \cmidrule{2-6} 
 & \textbf{\makecell{FT \\ (Audio)}} & 36.49 & 42.86 & 46.05 & 27.40 \\ \cmidrule{2-6} 
 & \textbf{KE} & 14.86 & 20.78 & 22.37 & 19.18 \\ \cmidrule{2-6} 
 & \textbf{MEND} & 17.57 & 18.18 & 19.74 & 17.81 \\ \cmidrule{2-6} 
 & \textbf{UnKE} & 12.16 & 16.88 & 21.05 & 20.55 \\ \cmidrule{2-6} 
 & \textbf{I-IKE} & 16.22 & 25.97 & 27.63 & 20.55 \\ \cmidrule{2-6} 
 & \textbf{IE-IKE} & 24.32 & 23.38 & 25.00 & 30.14 \\ \cmidrule{2-6} 
 & \textbf{WISE} & 12.16 & 20.78 & 22.37 & 17.81 \\
 \bottomrule
\end{tabular}%

\end{table}

%% file: tables/portability_analysis/speaker_gender.tex
\begin{table}
\centering
\caption{Breakdown of portability performance across different knowledge categories for the \textit{Speaker Gender} attribute. (\%)}
\label{tab:port_gender}

\begin{tabular}{cc|ccccc}
\toprule
\textbf{Model} & \textbf{Method} & \textbf{History} & \textbf{Celebrity} & \textbf{Cloth} & \textbf{Title} & \textbf{Vocal} \\ \midrule

\multirow[c]{11.5}{*}{\textbf{DeSTA}}

 & \textbf{\makecell{FT \\ (LLM)}} & 6.35 & 2.99 & 3.33 & 6.56 & 12.24 \\ \cmidrule{2-7} 
 & \textbf{\makecell{FT \\ (Audio)}} & 85.71 & 80.60 & 71.67 & 93.44 & 81.63 \\ \cmidrule{2-7} 
 & \textbf{KE} & 7.94 & 4.48 & 10.00 & 4.92 & 12.24 \\ \cmidrule{2-7} 
 & \textbf{MEND} & 4.76 & 2.99 & 3.33 & 3.28 & 10.20 \\ \cmidrule{2-7} 
 & \textbf{UnKE} & 4.76 & 4.48 & 8.33 & 4.92 & 8.16 \\ \cmidrule{2-7} 
 & \textbf{I-IKE} & 80.95 & 53.73 & 70.00 & 78.69 & 83.67 \\ \cmidrule{2-7} 
 & \textbf{IE-IKE} & 36.51 & 23.88 & 18.33 & 45.90 & 40.82 \\ \cmidrule{2-7} 
 & \textbf{WISE} & 33.33 & 22.39 & 31.67 & 36.07 & 24.49 \\
\midrule

\multirow[c]{11.5}{*}{\textbf{Qwen}}

 & \textbf{\makecell{FT \\ (LLM)}} & 47.62 & 11.94 & 18.33 & 32.79 & 18.37 \\ \cmidrule{2-7} 
 & \textbf{\makecell{FT \\ (Audio)}} & 90.48 & 35.82 & 40.00 & 88.52 & 57.14 \\ \cmidrule{2-7} 
 & \textbf{KE} & 61.90 & 28.36 & 36.67 & 40.98 & 20.41 \\ \cmidrule{2-7} 
 & \textbf{MEND} & 63.49 & 16.42 & 25.00 & 52.46 & 20.41 \\ \cmidrule{2-7} 
 & \textbf{UnKE} & 57.14 & 32.84 & 36.67 & 34.43 & 26.53 \\ \cmidrule{2-7} 
 & \textbf{I-IKE} & 77.78 & 28.36 & 31.67 & 55.74 & 28.57 \\ \cmidrule{2-7} 
 & \textbf{IE-IKE} & 65.08 & 28.36 & 33.33 & 36.07 & 36.73 \\ \cmidrule{2-7} 
 & \textbf{WISE} & 68.25 & 22.39 & 30.00 & 59.02 & 22.45 \\
\midrule

\multirow[c]{11.5}{*}{\textbf{AF}}

 & \textbf{\makecell{FT \\ (LLM)}} & 76.19 & 40.30 & 51.67 & 49.18 & 38.78 \\ \cmidrule{2-7} 
 & \textbf{\makecell{FT \\ (Audio)}} & 82.54 & 43.28 & 55.00 & 81.97 & 53.06 \\ \cmidrule{2-7} 
 & \textbf{KE} & 76.19 & 41.79 & 48.33 & 49.18 & 40.82 \\ \cmidrule{2-7} 
 & \textbf{MEND} & 74.60 & 41.79 & 48.33 & 50.82 & 40.82 \\ \cmidrule{2-7} 
 & \textbf{UnKE} & 77.78 & 32.84 & 51.67 & 60.66 & 42.86 \\ \cmidrule{2-7} 
 & \textbf{I-IKE} & 73.02 & 44.78 & 60.00 & 65.57 & 46.94 \\ \cmidrule{2-7} 
 & \textbf{IE-IKE} & 77.78 & 44.78 & 60.00 & 67.21 & 57.14 \\ \cmidrule{2-7} 
 & \textbf{WISE} & 71.43 & 38.81 & 50.00 & 44.26 & 38.78 \\
 \bottomrule
\end{tabular}%

\end{table}

%% file: tables/portability_analysis/spoken_language.tex
\begin{table}
\centering
\caption{Breakdown of portability performance across different knowledge categories for the \textit{Spoken Language} attribute. (\%)}
\label{tab:port_language}

\resizebox{\textwidth}{!}{%
\begin{tabular}{cc|ccccccccc}
\toprule
\textbf{Model} & \textbf{Method} & \textbf{Iso} & \textbf{City} & \textbf{Dish} & \textbf{History} & \textbf{Language Family} & \textbf{Literary} & \textbf{Official Language} & \textbf{Test} & \textbf{Translation} \\ \midrule

\multirow[c]{11.5}{*}{\textbf{DeSTA}}

 & \textbf{\makecell{FT \\ (LLM)}} & 10.71 & 41.18 & 37.14 & 38.89 & 14.81 & 2.86 & 23.53 & 30.77 & 40.62 \\ \cmidrule{2-11} 
 & \textbf{\makecell{FT \\ (Audio)}} & 67.86 & 58.82 & 74.29 & 66.67 & 74.07 & 54.29 & 58.82 & 41.03 & 81.25 \\ \cmidrule{2-11} 
 & \textbf{KE} & 14.29 & 23.53 & 22.86 & 30.56 & 14.81 & 11.43 & 26.47 & 15.38 & 40.62 \\ \cmidrule{2-11} 
 & \textbf{MEND} & 14.29 & 23.53 & 28.57 & 22.22 & 11.11 & 14.29 & 5.88 & 41.03 & 46.88 \\ \cmidrule{2-11} 
 & \textbf{UnKE} & 32.14 & 17.65 & 5.71 & 16.67 & 18.52 & 8.57 & 5.88 & 23.08 & 28.12 \\ \cmidrule{2-11} 
 & \textbf{I-IKE} & 82.14 & 70.59 & 77.14 & 72.22 & 59.26 & 48.57 & 61.76 & 66.67 & 84.38 \\ \cmidrule{2-11} 
 & \textbf{IE-IKE} & 57.14 & 26.47 & 37.14 & 22.22 & 11.11 & 25.71 & 23.53 & 23.08 & 65.62 \\ \cmidrule{2-11} 
 & \textbf{WISE} & 78.57 & 20.59 & 17.14 & 2.78 & 37.04 & 2.86 & 41.18 & 12.82 & 34.38 \\
\midrule

\multirow[c]{11.5}{*}{\textbf{Qwen}}

 & \textbf{\makecell{FT \\ (LLM)}} & 3.57 & 55.88 & 34.29 & 36.11 & 3.70 & 5.71 & 26.47 & 33.33 & 50.00 \\ \cmidrule{2-11} 
 & \textbf{\makecell{FT \\ (Audio)}} & 75.00 & 44.12 & 40.00 & 38.89 & 44.44 & 31.43 & 64.71 & 28.21 & 62.50 \\ \cmidrule{2-11} 
 & \textbf{KE} & 3.57 & 26.47 & 28.57 & 19.44 & 14.81 & 17.14 & 14.71 & 30.77 & 46.88 \\ \cmidrule{2-11} 
 & \textbf{MEND} & 17.86 & 29.41 & 25.71 & 13.89 & 18.52 & 17.14 & 26.47 & 23.08 & 46.88 \\ \cmidrule{2-11} 
 & \textbf{UnKE} & 64.29 & 35.29 & 14.29 & 22.22 & 51.85 & 20.00 & 32.35 & 15.38 & 43.75 \\ \cmidrule{2-11} 
 & \textbf{I-IKE} & 10.71 & 35.29 & 20.00 & 19.44 & 18.52 & 14.29 & 17.65 & 23.08 & 46.88 \\ \cmidrule{2-11} 
 & \textbf{IE-IKE} & 7.14 & 26.47 & 25.71 & 13.89 & 29.63 & 14.29 & 8.82 & 25.64 & 50.00 \\ \cmidrule{2-11} 
 & \textbf{WISE} & 78.57 & 44.12 & 45.71 & 36.11 & 40.74 & 5.71 & 44.12 & 33.33 & 59.38 \\
\midrule

\multirow[c]{11.5}{*}{\textbf{AF}}

 & \textbf{\makecell{FT \\ (LLM)}} & 17.86 & 20.59 & 14.29 & 25.00 & 29.63 & 34.29 & 32.35 & 33.33 & 37.50 \\ \cmidrule{2-11} 
 & \textbf{\makecell{FT \\ (Audio)}} & 92.86 & 32.35 & 51.43 & 55.56 & 55.56 & 31.43 & 41.18 & 35.90 & 46.88 \\ \cmidrule{2-11} 
 & \textbf{KE} & 28.57 & 20.59 & 25.71 & 25.00 & 37.04 & 31.43 & 17.65 & 30.77 & 37.50 \\ \cmidrule{2-11} 
 & \textbf{MEND} & 21.43 & 20.59 & 17.14 & 27.78 & 25.93 & 34.29 & 20.59 & 33.33 & 40.62 \\ \cmidrule{2-11} 
 & \textbf{UnKE} & 10.71 & 20.59 & 17.14 & 22.22 & 18.52 & 31.43 & 23.53 & 28.21 & 40.62 \\ \cmidrule{2-11} 
 & \textbf{I-IKE} & 25.00 & 26.47 & 25.71 & 25.00 & 29.63 & 37.14 & 23.53 & 30.77 & 37.50 \\ \cmidrule{2-11} 
 & \textbf{IE-IKE} & 57.14 & 32.35 & 25.71 & 33.33 & 33.33 & 34.29 & 35.29 & 25.64 & 40.62 \\ \cmidrule{2-11} 
 & \textbf{WISE} & 42.86 & 23.53 & 22.86 & 25.00 & 25.93 & 25.71 & 26.47 & 25.64 & 37.50 \\
 \bottomrule
\end{tabular}%
}

\end{table}